\documentclass[reprint,amsmath,amssymb,aps,superscriptaddress,nofootinbib]{revtex4-2}
\usepackage[pdftex]{graphicx}
\usepackage{dcolumn}
\usepackage{bm}
\usepackage{float}
\usepackage{hyperref}
\usepackage{epstopdf}
\usepackage{mathrsfs}
\usepackage{graphicx}
\usepackage{multirow}
\usepackage{amsmath}
\usepackage{amssymb}
\usepackage{xcolor}

\definecolor{myGreen}{HTML}{D1E7DD}
\definecolor{myBlue}{HTML}{CFE2F3}

\newcommand{\Fig}[1]{Fig.\,\ref{#1}}
\newcommand{\Eq}[1]{Eq.\,\eqref{#1}}

\hypersetup{
            colorlinks=true,
            linkcolor=blue,
            anchorcolor=blue,
            citecolor=blue}

\definecolor{myGreen}{HTML}{D1E7DD}
\definecolor{myBlue}{HTML}{CFE2F3}
\definecolor{termColor}{HTML}{FFF2CC}

\begin{document}

\title{Intrinsic violation of the Wiedemann-Franz law in interacting systems}

\author{YuanDong Wang}
\email{ydwang@cau.edu.cn}
\affiliation{
Department of Applied Physics, College of Science, China Agricultural University, Qinghua East Road, Beijing 100083, China}

\author{Zhen-Gang Zhu}
\email{zgzhu@ucas.ac.cn}
\affiliation{
School of Electronic, Electrical and Communication Engineering, University of Chinese Academy of Sciences, Beijing 100049, China. \\
 }


\begin{abstract}

The Wiedemann-Franz (WF) law dictates a universal ratio
between thermal and electrical conductivities, is widely obeyed by  Fermi liquid systems. Here, we identify a fundamental yet often overlooked, thermodynamic mechanism for the violation of WF law:  the temperature-dependent renormalization of the electronic band structure. We demonstrate that the interaction-induced energy drift $\partial\epsilon_k/\partial T$, acts as an effective driving force that fundamentally decouples heat transport from charge transport. We derive a generalized transport relation linking the Lorenz ratio deviation directly to the thermoelectric response.  Our findings provide a unified framework for understanding thermal transport in interacting topological phases and suggest the Lorenz ratio as a probe for distinguishing topological robustness from Fermi liquid instabilities.

\end{abstract}

\maketitle

The Wiedemann-Franz (WF) law constitutes one of the fundamental pillars of condensed matter physics, asserting a universal ratio between thermal conductivity $\kappa$ and electrical conductivity $\sigma$ of a metal \cite{wflaw}.
The universal ratio is the famous Lorenz number $L_0 = \kappa / (\sigma T)=(\pi^2/3)(k_B/e)^2$ (Sommerfeld value), independent of material details.
The validity of the WF law relies on the basic assumption of Fermi liquid theory, where charge and heat are carried by the same quasiparticles, and both transport processes are governed by the same relaxation time.
Numbers of experiments have verified the validity of this law in the zero-temperature limit \cite{PhysRevLett.101.046401,PhysRevLett.110.236402,PhysRevB.89.045130,PhysRevLett.115.046402}.

However, deviations from the WF law are observed and serve as a diagnostic tool for exotic scattering mechanisms or non-Fermi liquid behaviors.
Conventionally, the downward deviation of the Lorenz ratio ($L < L_0$) at finite temperatures is attributed to inelastic scatterings.
In simple metals, this arises from the vertical scattering of electrons by phonons (small-angle scattering), which degrades the energy current more efficiently than the charge current  \cite{ziman2001electrons, mahan2013many}.
In strongly correlated systems, such as heavy fermions, similar violations are often driven by inelastic scattering off spin fluctuations near magnetic instabilities \cite{PhysRevLett.73.3294, PhysRevLett.94.216602, doi:10.1126/science.1140762, PhysRevLett.102.156404, Pfau2012} and fermionic quantum critical fluctuations \cite{cpl_41_12_127401}.
Violations from WF law have been reported in charge-density-wave material associated with critical fluctuations \cite{PhysRevB.104.L241109}, as well as in topological semimetals \cite{Gooth2018,Jaoui2018, Wang2025, zhong2025phonon}.
Furthermore, in the hydrodynamic regime, dominant electron-electron inelastic collisions with the kinematic decoupling of charge and heat relaxation can also lead to a deviation from $L_0$ \cite{PhysRevLett.115.056603,PhysRevLett.115.046402,doi:10.1126/science.aad0343, PhysRevB.88.125107, PhysRevB.97.245128, PhysRevB.98.115130,Zarenia_2019,Jaoui2021,annurev, Hui_2025}.

In view of the versatile results in the research on WF law, a fundamental issue yet unanswered question is whether WF law violation occurs occasionally, or is it ubiquitous in principle?
To explore this issue, we consider the effect of electron-electron (ee) interactions in the framework of standard Fermi liquid theory.
This choice is made by considering the universality of ee interactions in realistic materials. 
A remarkable find is that WF law is generally not satisfied when many-body interaction is taken into account. This stems from a simple but profound fact that the band energy $\epsilon_k$ gains a temperature dependence from the renormalization of ee interaction. 
This effect is usually ignored since it only manifests itself at finite temperature rather than ground state (zero temperature).  
Technically speaking, this renormalization generally exists by mean-field method, Dyson equation for selfenergies, and infinite-order Feynman diagrams. It may not be present for finite-order Feynman-diagram calculation.
The gains of temperature dependence for the band energy, i.e. $\partial\epsilon_k/\partial T \neq 0$, bring us a profound consequence. 
%
It generates an internal thermodynamic force that selectively drives the entropy flow (heat current) but does not affect the conserved charge current, inherently breaking the parallelism required for the WF law. 

In this letter, we present a systematic derivation and show that the interaction-induced energy drift (IED) $\partial \epsilon_{\bm{k}} / \partial T$ leads to a correction in the thermal conductivity that violates the WF law.
For longitudinal transport, we demonstrate that this deviation is proportional to the Seebeck coefficient, implying that systems with strong thermoelectric response will inherently exhibit large deviations from $L_0$. 
Furthermore, we extend our theory to the transverse (Hall) channel. We reveal that the violation of the transverse WF law is governed by the transverse Seebeck (anomalous Nernst) coefficient and the Berry curvature density. 
In particular, we predict that in the quantum anomalous Hall (QAH) regime, where the transport is topologically quantized, the WF law is strictly protected despite the presence of interactions, providing a sharp distinction from the metallic regime.

For the charge current, $\mathbf{J} = \hat{\sigma}\mathbf{E}$, the conductivity tensor can be written in form of  $\hat{\sigma} = \int \hat{\Sigma}(\varepsilon) \left(-\frac{\partial f_0}{\partial \varepsilon}\right) d\varepsilon $, with $\Sigma_{\alpha\beta}(\epsilon)$ being the transport kernel. 
The longitudinal transport depends on the density of states $N(\epsilon)$ and group velocity $v_x$, i.e. $\Sigma_{xx} = e^2 \tau(\epsilon) v_x^2 N(\epsilon)$; 
while for the anomalous Hall response, it is determined by the integrated Berry curvature, $\Sigma_{xy} = -(e^2/\hbar) \sum_n \int [d\mathbf{k}] \Omega_n^z \Theta(\epsilon - \epsilon_{n\mathbf{k}})$ \cite{SM}.

In standard Fermi liquid theory, the band structure is assumed to be rigid. However, in the presence of ee interactions, the quasiparticle dispersion $\epsilon_\mathbf{k}(T)$ acquires a temperature dependence due to the renormalization of the mean-field potential. In particular, the variation of $\epsilon_{\bm{k}}(T)$ could be drastic near the phase transition point.
This spectral shift modifies the driving force for the heat current. The linearized Boltzmann equation involves the gradient of the distribution function $f_0$:
\begin{equation}
    \nabla f_0 = \frac{\partial f_0}{\partial \epsilon} \left( \nabla \mu + \frac{\epsilon-\mu}{T}\nabla T - \frac{\partial \epsilon_\mathbf{k}}{\partial T} \nabla T \right).
    \label{eq:grad_f0}
\end{equation}
The last term in Eq.~(\ref{eq:grad_f0}) represents an interaction-induced effective force, $\mathbf{F}_{\text{int}} \propto \nabla T (\partial \epsilon_\mathbf{k}/\partial T)$, which drives the entropy flow but is absent in the charge transport equation.

Substituting Eq.~(\ref{eq:grad_f0}) into the definition of the heat current $\mathbf{J}_Q  =  - \hat{\kappa}\nabla T = \int d\epsilon (\epsilon-\mu) \mathbf{v}_\mathbf{k} \delta f$, and performing the Sommerfeld expansion up to $O(T^2)$, the thermal conductivity tensor splits into two contributions: $\hat{\kappa} = \hat{\kappa}_0 + \hat{\kappa}_{\text{int}}$.
The first term recovers the conventional WF law, $\hat{\kappa}_0 = L_0 T \hat{\sigma}$. 
The second term arises entirely from the IED as 
\begin{equation}
    \hat{\kappa}_{\text{int}} \approx -\frac{\pi^2 k_B^2 T^2}{3e^2} \left( \frac{\partial \hat{\Sigma}}{\partial \epsilon} \right)_\mu \left( \frac{\partial \epsilon}{\partial T} \right)_\mu.
\end{equation}
Combining these terms, the Lorenz ratio $L = \kappa/(\sigma T)$ deviates  from the standard Sommerfeld value $L_0$.  The relative deviation, quantified by the factor $\hat{\gamma} = L/L_0$, takes the compact form (More details in ~\cite{SM}):
\begin{equation}
    \hat{\gamma}(T) = 1 - \frac{1}{L_0 e} \hat{\mathcal{S}} \left( \frac{\partial \epsilon}{\partial T} \right)_\mu.
    \label{eq:gamma_final}
\end{equation}
For convenience, we exploit $\gamma_{xx(xy)}-1$ to indicate the deviation from the WF law, and call it as longitudinal (transverse) Wiedemann-Franz deviation where $\gamma_{xx(xy)}$ reads the longitudinal (transverse) component of $\hat{\gamma}$ tensor.
Remarkably, it is natural to introduce a new physical quantity named as \textit{Mott ratio tensor} 
$\mathcal{S}_{\alpha\beta}=eL_{0}T\frac{1}{\sigma_{\alpha\beta}(\mu)}\left(\frac{\partial {\sigma}_{\alpha\beta}}{\partial \varepsilon}\right)_\mu$.
By use of the conventional Mott relation, it can be rewritten by the thermoelectric conductivity $\alpha_{\alpha\beta}$ as $\mathcal{S}_{\alpha\beta}= \alpha_{\alpha\beta}/\sigma_{\alpha\beta}$,  which is closely related to the macroscopic Seebeck coefficients $S_{\alpha\beta}$ measured under open-circuit conditions.   The longitudinal component $\mathcal{S}_{xx} = \alpha_{xx}/\sigma_{xx} \approx S_{xx}$ serves as a valid approximation for metals owing to small Hall angle \cite{SM}. Conversely, the transverse (anomalous) component $\mathcal{S}_{xy} = \alpha_{xy}/\sigma_{xy}$ defines the \textit{anomalous Mott ratio}, which is related to the anomalous Nernst effect. This ratio fundamentally governs the transverse Wiedemann-Franz law deviation.

Eq.~(\ref{eq:gamma_final}) is our key result. It reveals that the violation of the WF law is governed by the product of the thermoelectric response ($\hat{S}$) and the IED ($\partial \epsilon / \partial T$). 
For transverse transport, $\hat{S}$ becomes the anomalous Nernst coefficient, which is proportional to the energy derivative of the anomalous Hall conductivity, $\partial \sigma_{xy}/\partial \epsilon$. 
Consequently, in the Quantum Anomalous Hall (QAH) regime where $\sigma_{xy}$ is quantized to a topological integer, the term $\partial \hat{\Sigma}_{xy}/\partial \epsilon$ vanishes. 
This dictates that $\gamma_{xy} \to 1$, implying that the transverse WF law is topologically protected against interaction-induced spectral shifts, providing a robust experimental signature.

\begin{figure}[htbp]
\centering
\includegraphics [width=\columnwidth]{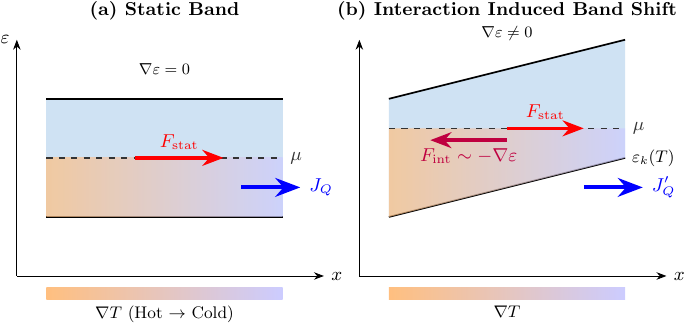}
\caption{Schematic of the interaction-induced force. The orange-to-blue gradient represents the thermal excitation density, fully covering the band below the Fermi level. (a) Rigid band: The statistical force $F_{\text{stat}}$ drives transport. (b) Interacting system: The band tilt $\nabla \varepsilon$ generates an opposing interaction force $F_{\text{int}}$, modifying the heat current.}\label{fig1}
\end{figure}

To elucidate the physical origin of the Wiedemann-Franz law violation, we contrast the transport mechanisms in rigid and fluctuating band structures, as illustrated in \Fig{fig1}.
In the conventional non-interacting picture \Fig{fig1}(a), quasiparticles traverse a static energy landscape. Both charge and heat currents are driven by the same statistical forces-gradients in the chemical potential and temperature-acting on the Fermi distribution. Consequently, the ratio of thermal to electrical conductivity is fixed by fundamental constants, yielding the universal Lorenz number $L_0$.

However, ee interactions fundamentally alter this scenario by making the band structure temperature-dependent, i.e., $\epsilon_{\mathbf{k}} \to \epsilon_{\mathbf{k}}(T)$. 
As depicted in \Fig{fig1}(b), a temperature gradient $\nabla T$ applied across the sample not only creates a statistical distribution imbalance but also induces a spatial variation in the local band energy, $\nabla \epsilon = (\partial \epsilon_{\mathbf{k}}/\partial T)\nabla T$. 
Thermodynamically, this spectral shift manifests as an additional, interaction-induced driving force $\mathbf{F}_{\text{int}}$. 
Since the heat current represents the flow of entropy (energy relative to the chemical potential), it is sensitive to this redistribution of energy levels, whereas the charge conservation protects the electrical current from such spectral drifts. 
This decoupling, driven by the variation of the band structure shown in \Fig{fig1}(b), is the key mechanism responsible for the failure of WF law.

To explicitly demonstrate the violation of WF law arising from band renormalization, we apply our theory to the honeycomb lattice subject to Rashba SOC and an exchange field. This system is particularly suitable for our study because it hosts diverse topological phase transitions driven by the interplay between spin-orbit coupling and magnetic ordering \cite{PhysRevB.82.161414,PhysRevB.82.075106,PhysRevB.83.155447}. The tight-binding Hamiltonian is
\begin{eqnarray}
H_0 &=& -t\sum_{\langle i,j \rangle}\sum_{\sigma}c_{i\sigma}^\dagger c_{j\sigma} + i \lambda_R \sum_{\langle i,j \rangle, \alpha \beta} c_{i\alpha}^\dagger (\boldsymbol{\sigma} \times \mathbf{d}_{ij})_z c_{j\beta}  \notag\\
&+& M \sum_{i, \sigma \sigma'} c_{i\sigma}^\dagger \sigma^z_{\sigma \sigma'} c_{i\sigma'} -\mu\sum_{i}\sum_{\sigma}c_{i\sigma}^\dagger c_{i\sigma}.
\label{Hamiltonian0}
\end{eqnarray}
Here $c_i$ is an electron annihilation operator either on sublattice A or B. $t$ is the hopping integral, $\langle i,j \rangle$ denotes nearest-neighbor. 
The second term represents the Rashba SOC with strength $\lambda_R$, opens a hybridization gap when bands with opposite spins cross. $\mathbf{d}_{ij}$ represents a unit vector pointing from site $j$ to site $i$.
The third term encapsulates the symmetry-breaking potentials: $M$ is the proximity-induced ferromagnetic exchange field arising from the magnetic substrate, acting as a uniform Zeeman field.  The final term is the chemical potential.

\begin{figure}[tb]
\centering
\includegraphics [width=\columnwidth]{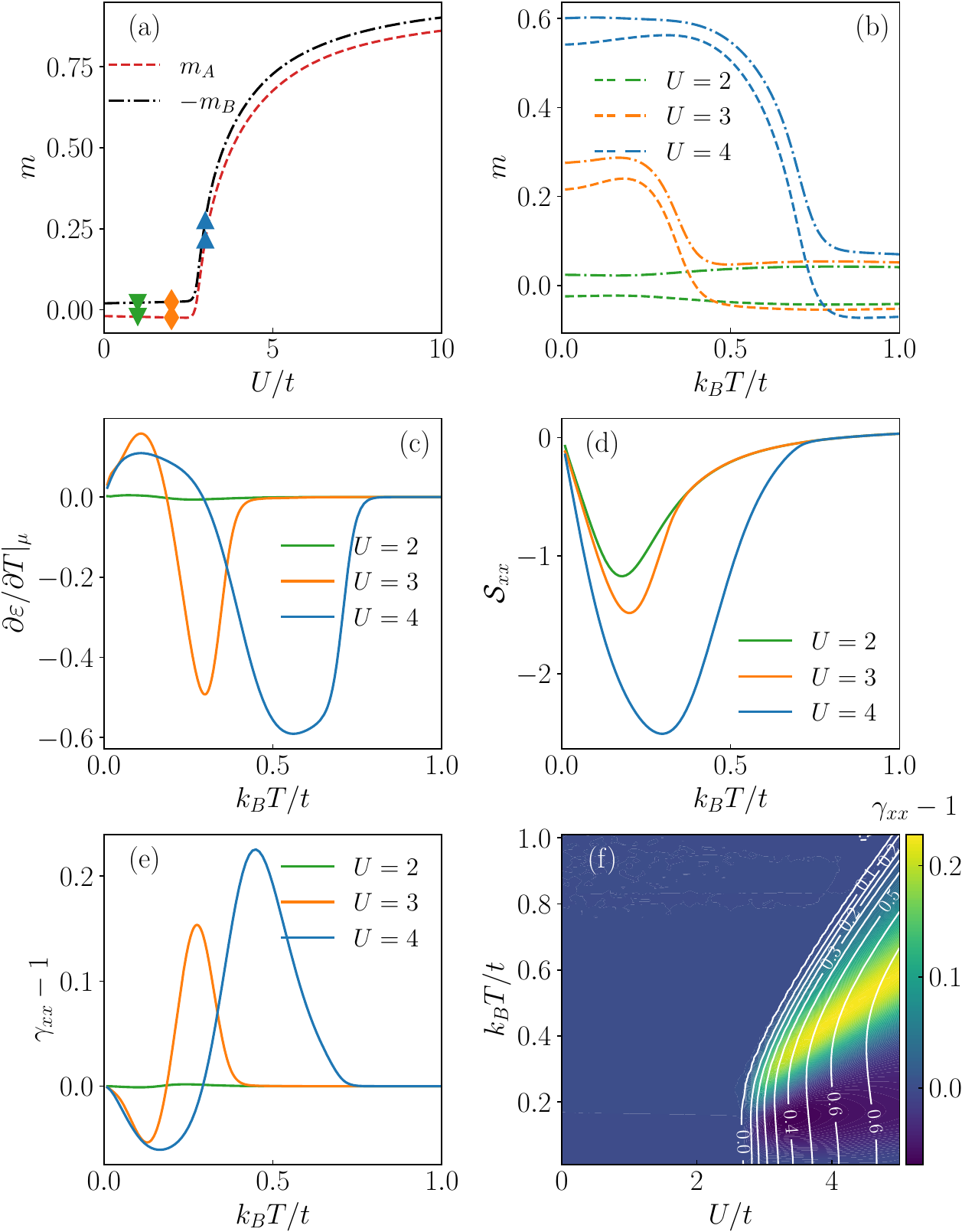}
\caption{
(a) Sublattice magnetization $m_A$ and $-m_B$ versus Hubbard interaction $U$ at zero temperature. (b)-(e) Sublattice magnetization, interaction-induced energy drift, longitudinal Mott ratio, and the longitudinal WF law deviation $\gamma_{xx}(T) -1$ as a function of temperature for different $U$. (f) A color density plot showing $\gamma_{xx}-1$, in the U-T plane. The magnitude of $\gamma_{xx}-1$ is represented by the color density, which is superimposed with white contour lines denoting the sublattice magnetization $m$.  The parameters are $t=1$, $\lambda_R = 0.1$, $M=0.1$, and particle filling number is set to $2.1$.
}\label{fig2}
\end{figure}

To explore the effects of ee correlations on the topological transport, we include the on-site repulsive Hubbard interaction
\begin{equation}
    H_{\text{int}} = U \sum_{i} n_{i\uparrow} n_{i\downarrow},
\end{equation}
where $n_{i\sigma} = c_{i\sigma}^\dagger c_{i\sigma}$ is the number operator. In the limit of intermediate interaction strength, we treat this term within the Hartree-Fock mean-field approximation. We decompose the interaction term by introducing the local magnetization order parameter $\langle m_i \rangle = \langle n_{i\uparrow} \rangle - \langle n_{i\downarrow} \rangle$ and the local charge density $\langle n_i \rangle = \langle n_{i\uparrow} \rangle + \langle n_{i\downarrow} \rangle$. Neglecting the fluctuation terms (e.g., $(n_{i\uparrow} - \langle n_{i\uparrow} \rangle)(n_{i\downarrow} - \langle n_{i\downarrow} \rangle)$) and assuming a uniform charge distribution, the interacting Hamiltonian simplifies to a quadratic form:
\begin{equation}
    H_{\text{int}}^{\text{MF}} = \sum_{i} \left[ -\frac{U}{2} \langle m_i \rangle (c_{i\uparrow}^\dagger c_{i\uparrow} - c_{i\downarrow}^\dagger c_{i\downarrow}) + \frac{U}{4} \langle m_i \rangle^2 \right].
\end{equation}
This effective potential acts as a local magnetic field dependent on the site index $i$ (for details see SM \cite{SM}).

The spin-orbit coupling opens a gap at the Dirac points ($K$ and $K^\prime$). For half-filling, a large $U$ implies a Mott transition between paramagnetic phase and SDW phase \cite{S_Sorella_1992, Jafari2009, PhysRevB.82.075106}. The sublattice magnetization versus $U$ at zero temperature is shown in \Fig{fig2}(a), where the paramagnetic-SDW phase transition occurs at $U\sim 2.7$. Therefore, for  $U>2.7$, the SDW-paramagnetic phase transition is expected by increasing temperature.
As we already know, a large WF deviation is anticipated near the phase transition point. The magnetization $m$ as a function of temperature $k_B T$ with varying $U$ is shown in \Fig{fig2}(b). The different values of $U$ are marked by same colors in \Fig{fig2}(a).  For the $U=4$ and $3$ which belong to SDW phase, SDW-paramagnetic phase transition occurs by increasing temperature, in contrast to $U=2$ for which the interaction is weak with a paramagnetic phase at zero temperature.

To investigate transport properties, the system is doped away from half-filling (by setting filling number $n=2.1$) to ensure a metal state.
In \Fig{fig2}(c) we depict the IED as a function of temperature. For $U=4$ and $3$, the magnitudes of IED are maximized around the phase transition points. However, that of $U=2$ is near to zero, owing to the absence of phase transition.
The deviation of WF law along the longitudinal direction, namely, $\gamma_{xx}(T)-1$, is depicted in \Fig{fig2}(e). The peaks of the IED of $U=4$ and $3$ locate at the phase transition points in \Fig{fig2}(b),  where we see that the large deviation of transverse WF law  is prominent with  drastic changes of IED around the phase transition points.
Noting that the peaks of $\gamma_{xx}(T)-1$ are not exactly coincident with that of IED, owing to the  existence of the Mott ratio in \Eq{eq:gamma_final}, shown in \Fig{fig2}(d). In \Fig{fig2}(f) we show the deviation of the longitudinal Wiedemann-Franz law, $\gamma_{xx} - 1$, in the $U$-$T$ plane.
The color density plot represents the magnitude of the $\gamma_{xx} - 1$, superimposed with white contour lines denoting the sublattice magnetization $m$. The maximal violation (bright regions) coincides perfectly with the dense contour lines, indicating that the breakdown of the WF law is most pronounced near the thermal phase transition boundary.

\begin{figure}[tb]
\centering
\includegraphics [width=\columnwidth]{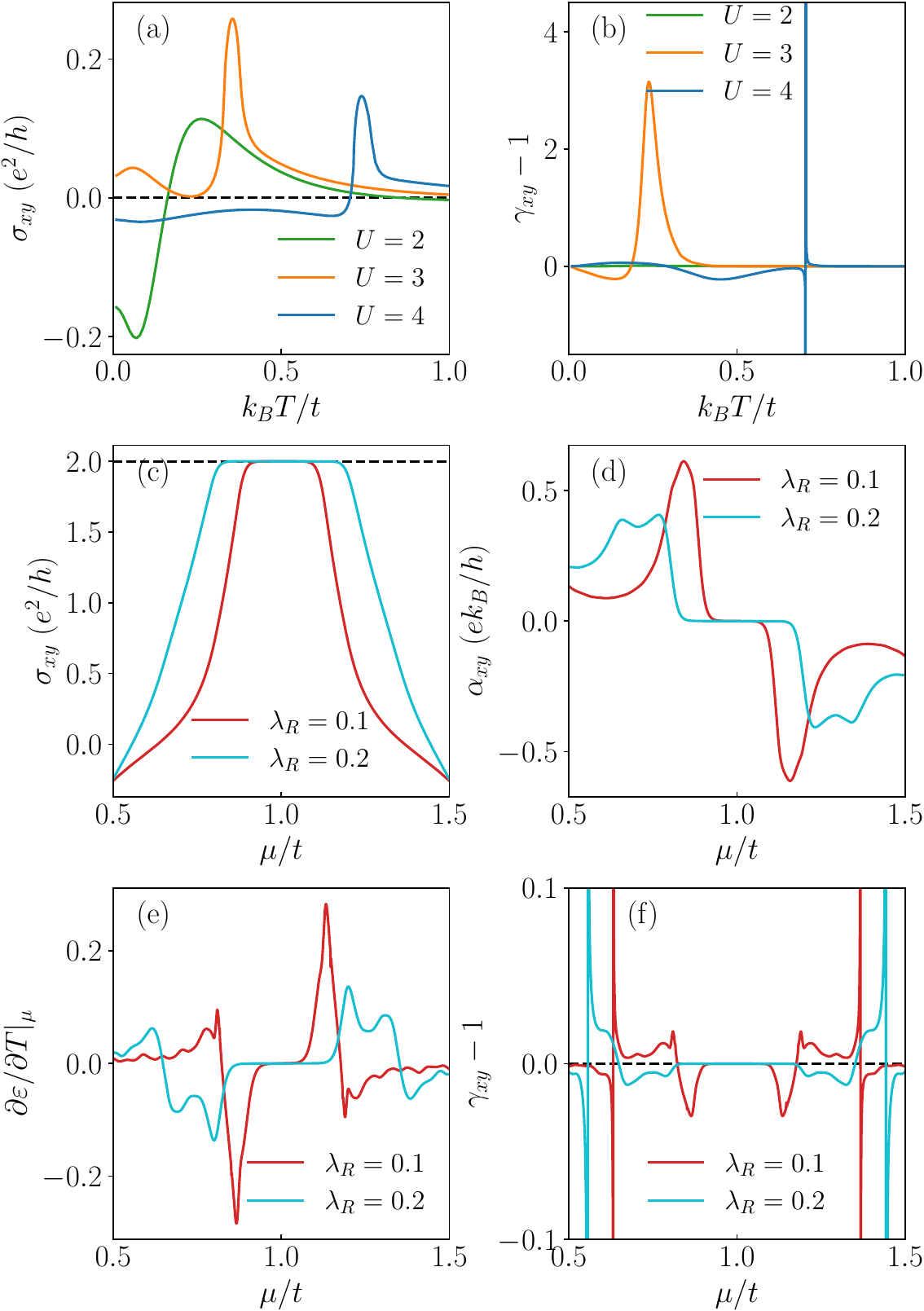}
\caption{(a) Anomalous Hall conductivity and (b) transverse WF deviation versus temperature for a metal state with filling number $n=2.1$. (c)-(f)  Topological protection of the transverse Wiedemann-Franz (WF) law against electron-electron interactions.
Dependence of transport quantities on the chemical potential $\mu/t$ for Rashba spin-orbit coupling strengths $\lambda_R=0.1$  and $\lambda_R=0.2$.
(c) Transverse electrical conductivity $\sigma_{xy}$.
(d) Transverse thermoelectric conductivity $\alpha_{xy}$.
(e) The interaction-induced energy drift at the Fermi level, $(\partial\epsilon/\partial T)_{\mu}$.
(f) The relative deviation of the transverse Lorenz ratio, $\gamma_{xy}-1$.
}\label{fig3}
\end{figure}

Next we study the possibility of the WF law deviation in transverse direction. To obtain Hall signal  $M=0.1$ is introduced to break the time-reversal symmetry. The violation is proportional to the anomalous (transverse) Mott ratio, $\alpha_{xy}/\sigma_{xy}$, which depends on the energy derivative of the integrated Berry curvature.
Remarkably, referring to \Eq{eq:gamma_final}, the violation diverges when Hall electrical conductivity approaches zero. In contrast, the divergence in the longitudinal situation  is absent since the longitudinal electrical conductivity is always finite. The Hall conductivity as a function of temperature is depicted in \Fig{fig3}(a).
The transverse  WF deviation $\gamma_{xy} -1$ as a function of temperature is shown in \Fig{fig3}(b).
Accordingly, the peak of $\gamma_{xy} -1$ of $U=2$ is attributed to $\sigma_{xy}\rightarrow 0$.
For the same reason, the zero point of $\sigma_{xy}$ of $U=4$ leads to the divergence of $\gamma_{xy} -1$. 
While previous studies attributed the violation of the transverse WF law to the mismatch between the thermal and electrical weighting of the Berry curvature \cite{doi:10.1126/sciadv.aaz3522,Ding_2021,PhysRevB.105.205104}, or to the the oscillatory channels of correlated metals due to mesoscopic interference effects \cite{Zhang2024}, our work highlights a distinct interaction-induced thermodynamic pathway.

Finally, we fix the temperature and investigate the dependence of the transverse transport properties on the chemical potential $\mu$, as illustrated in \Fig{fig3}(c)-(d).
$\sigma_{xy}$  develops a distinct quantized plateau of $2e^2/h$ when $\mu$ lies in the energy gap, indicating that the system is in the quantum anomalous Hall (QAH) phase with a Chern number of $\mathcal{C}=2$. 
When $\mu$ lies within a global bulk energy gap (the QAH regime), there are no bulk states at the Fermi level.
In this topological regime, $\sigma_{xy}$ is  independent of energy variation, i.e., $\partial\Sigma^{xy}/\partial\epsilon = 0$.
Consequently, the transverse thermoelectric conductivity $\alpha_{xy}$, shown in \Fig{fig3}(d), vanishes identically within the gap, consistent with the transverse Mott relation.
Although the IED remains finite near the band edges [\Fig{fig3}(e)], the vanishing thermoelectric response dominates the transport behavior.
As a result, the relative deviation of the Lorenz ratio, $\gamma_{xy}-1$, depicted in \Fig{fig3}(f), drops strictly to zero within the topological gap.
This confirms that the validity of the transverse Wiedemann-Franz law is topologically protected in the QAH regime, in sharp contrast to the metallic regions (for example, $\mu < 0.8t$ or $\mu > 1.2t$ for $\lambda_{R}=0.1$) where the finite Fermi surface and strong ee interactions lead to significant violations.

In this work, we have presented a comprehensive theoretical framework to investigate the violation of the Wiedemann-Franz (WF) law in interacting electron systems. Beyond the conventional mechanisms involving inelastic scattering, we identified a fundamental source of violation arising from the breakdown of the rigid band approximation. Within the mean-field Hartree-Fock description, the renormalization of the electronic potential introduces a interaction-induced energy drift $\partial\epsilon_{\bm{k}}/\partial T$. Thermodynamically, this term acts as an additional driving force for the heat current, thereby decoupling thermal transport from charge transport.

Our analytical derivation establishes a direct quantitative link between the Lorenz ratio deviation and the thermoelectric properties of the material. For longitudinal transport in metals, we showed that the violation is proportional to the Seebeck coefficient and the interaction-induced energy drift. Consequently, systems exhibiting strong thermoelectric responses are inherently prone to violating the WF law. Extending this formulation to the transverse channel, we found that the anomalous thermal Hall conductivity acquires a correction governed by the transverse Mott ratio and the energy derivative of the integrated Berry curvature.
Applying our theory to a honeycomb lattice with Rashba SOC and exchange field, we numerically verified these predictions. We observed significant deviations from the Sommerfeld value $L_0$ in the vicinity of magnetic phase transitions (e.g., between paramagnetic and spin-density wave phases), where the electronic structure undergoes rapid thermal reconstruction.

Most notably, our study reveals a profound distinction between metallic and topological regimes in transverse transport. In the Quantum Anomalous Hall (QAH) regime, the quantization of the electrical Hall conductivity within the topological gap implies a vanishing transverse Mott ratio ($\mathcal{S}^{xy} \to 0$). As a result, the transverse WF law remains strictly protected and valid, even in the presence of strong electron-electron interactions. This topological protection suggests that the Lorenz ratio can serve as a sensitive experimental probe to distinguish between interaction-driven Fermi liquid instabilities and robust topological phases. Our findings provide a unified perspective on thermal transport in interacting topological materials, paving the way for further explorations into non-Fermi liquid thermal hydrodynamics.

This work is supported by the National Key R\&D Program of China (Grant No. 2024YFA1409200), NSFC (Grants No. 12404147), CAS Project for Young Scientists in Basic Research Grant No. YSBR-057.

\bibliographystyle{apsrev4-2}

\begin{thebibliography}{39}%
\makeatletter
\providecommand \@ifxundefined [1]{%
 \@ifx{#1\undefined}
}%
\providecommand \@ifnum [1]{%
 \ifnum #1\expandafter \@firstoftwo
 \else \expandafter \@secondoftwo
 \fi
}%
\providecommand \@ifx [1]{%
 \ifx #1\expandafter \@firstoftwo
 \else \expandafter \@secondoftwo
 \fi
}%
\providecommand \natexlab [1]{#1}%
\providecommand \enquote  [1]{``#1''}%
\providecommand \bibnamefont  [1]{#1}%
\providecommand \bibfnamefont [1]{#1}%
\providecommand \citenamefont [1]{#1}%
\providecommand \href@noop [0]{\@secondoftwo}%
\providecommand \href [0]{\begingroup \@sanitize@url \@href}%
\providecommand \@href[1]{\@@startlink{#1}\@@href}%
\providecommand \@@href[1]{\endgroup#1\@@endlink}%
\providecommand \@sanitize@url [0]{\catcode `\\12\catcode `\$12\catcode
  `\&12\catcode `\#12\catcode `\^12\catcode `\_12\catcode `\%12\relax}%
\providecommand \@@startlink[1]{}%
\providecommand \@@endlink[0]{}%
\providecommand \url  [0]{\begingroup\@sanitize@url \@url }%
\providecommand \@url [1]{\endgroup\@href {#1}{\urlprefix }}%
\providecommand \urlprefix  [0]{URL }%
\providecommand \Eprint [0]{\href }%
\providecommand \doibase [0]{https://doi.org/}%
\providecommand \selectlanguage [0]{\@gobble}%
\providecommand \bibinfo  [0]{\@secondoftwo}%
\providecommand \bibfield  [0]{\@secondoftwo}%
\providecommand \translation [1]{[#1]}%
\providecommand \BibitemOpen [0]{}%
\providecommand \bibitemStop [0]{}%
\providecommand \bibitemNoStop [0]{.\EOS\space}%
\providecommand \EOS [0]{\spacefactor3000\relax}%
\providecommand \BibitemShut  [1]{\csname bibitem#1\endcsname}%
\let\auto@bib@innerbib\@empty
\bibitem [{\citenamefont {Franz}\ and\ \citenamefont
  {Wiedemann}(1853)}]{wflaw}%
  \BibitemOpen
  \bibfield  {author} {\bibinfo {author} {\bibfnamefont {R.}~\bibnamefont
  {Franz}}\ and\ \bibinfo {author} {\bibfnamefont {G.}~\bibnamefont
  {Wiedemann}},\ }\href
  {https://doi.org/https://doi.org/10.1002/andp.18531650802} {\bibfield
  {journal} {\bibinfo  {journal} {Annalen der Physik}\ }\textbf {\bibinfo
  {volume} {165}},\ \bibinfo {pages} {497} (\bibinfo {year}
  {1853})}\BibitemShut {NoStop}%
\bibitem [{\citenamefont {Seyfarth}\ \emph {et~al.}(2008)\citenamefont
  {Seyfarth}, \citenamefont {Brison}, \citenamefont {Knebel}, \citenamefont
  {Aoki}, \citenamefont {Lapertot},\ and\ \citenamefont
  {Flouquet}}]{PhysRevLett.101.046401}%
  \BibitemOpen
  \bibfield  {author} {\bibinfo {author} {\bibfnamefont {G.}~\bibnamefont
  {Seyfarth}}, \bibinfo {author} {\bibfnamefont {J.~P.}\ \bibnamefont
  {Brison}}, \bibinfo {author} {\bibfnamefont {G.}~\bibnamefont {Knebel}},
  \bibinfo {author} {\bibfnamefont {D.}~\bibnamefont {Aoki}}, \bibinfo {author}
  {\bibfnamefont {G.}~\bibnamefont {Lapertot}},\ and\ \bibinfo {author}
  {\bibfnamefont {J.}~\bibnamefont {Flouquet}},\ }\href
  {https://doi.org/10.1103/PhysRevLett.101.046401} {\bibfield  {journal}
  {\bibinfo  {journal} {Phys. Rev. Lett.}\ }\textbf {\bibinfo {volume} {101}},\
  \bibinfo {pages} {046401} (\bibinfo {year} {2008})}\BibitemShut {NoStop}%
\bibitem [{\citenamefont {Machida}\ \emph {et~al.}(2013)\citenamefont
  {Machida}, \citenamefont {Tomokuni}, \citenamefont {Izawa}, \citenamefont
  {Lapertot}, \citenamefont {Knebel}, \citenamefont {Brison},\ and\
  \citenamefont {Flouquet}}]{PhysRevLett.110.236402}%
  \BibitemOpen
  \bibfield  {author} {\bibinfo {author} {\bibfnamefont {Y.}~\bibnamefont
  {Machida}}, \bibinfo {author} {\bibfnamefont {K.}~\bibnamefont {Tomokuni}},
  \bibinfo {author} {\bibfnamefont {K.}~\bibnamefont {Izawa}}, \bibinfo
  {author} {\bibfnamefont {G.}~\bibnamefont {Lapertot}}, \bibinfo {author}
  {\bibfnamefont {G.}~\bibnamefont {Knebel}}, \bibinfo {author} {\bibfnamefont
  {J.-P.}\ \bibnamefont {Brison}},\ and\ \bibinfo {author} {\bibfnamefont
  {J.}~\bibnamefont {Flouquet}},\ }\href
  {https://doi.org/10.1103/PhysRevLett.110.236402} {\bibfield  {journal}
  {\bibinfo  {journal} {Phys. Rev. Lett.}\ }\textbf {\bibinfo {volume} {110}},\
  \bibinfo {pages} {236402} (\bibinfo {year} {2013})}\BibitemShut {NoStop}%
\bibitem [{\citenamefont {Reid}\ \emph {et~al.}(2014)\citenamefont {Reid},
  \citenamefont {Tanatar}, \citenamefont {Daou}, \citenamefont {Hu},
  \citenamefont {Petrovic},\ and\ \citenamefont
  {Taillefer}}]{PhysRevB.89.045130}%
  \BibitemOpen
  \bibfield  {author} {\bibinfo {author} {\bibfnamefont {J.-P.}\ \bibnamefont
  {Reid}}, \bibinfo {author} {\bibfnamefont {M.~A.}\ \bibnamefont {Tanatar}},
  \bibinfo {author} {\bibfnamefont {R.}~\bibnamefont {Daou}}, \bibinfo {author}
  {\bibfnamefont {R.}~\bibnamefont {Hu}}, \bibinfo {author} {\bibfnamefont
  {C.}~\bibnamefont {Petrovic}},\ and\ \bibinfo {author} {\bibfnamefont
  {L.}~\bibnamefont {Taillefer}},\ }\href
  {https://doi.org/10.1103/PhysRevB.89.045130} {\bibfield  {journal} {\bibinfo
  {journal} {Phys. Rev. B}\ }\textbf {\bibinfo {volume} {89}},\ \bibinfo
  {pages} {045130} (\bibinfo {year} {2014})}\BibitemShut {NoStop}%
\bibitem [{\citenamefont {Taupin}\ \emph {et~al.}(2015)\citenamefont {Taupin},
  \citenamefont {Knebel}, \citenamefont {Matsuda}, \citenamefont {Lapertot},
  \citenamefont {Machida}, \citenamefont {Izawa}, \citenamefont {Brison},\ and\
  \citenamefont {Flouquet}}]{PhysRevLett.115.046402}%
  \BibitemOpen
  \bibfield  {author} {\bibinfo {author} {\bibfnamefont {M.}~\bibnamefont
  {Taupin}}, \bibinfo {author} {\bibfnamefont {G.}~\bibnamefont {Knebel}},
  \bibinfo {author} {\bibfnamefont {T.~D.}\ \bibnamefont {Matsuda}}, \bibinfo
  {author} {\bibfnamefont {G.}~\bibnamefont {Lapertot}}, \bibinfo {author}
  {\bibfnamefont {Y.}~\bibnamefont {Machida}}, \bibinfo {author} {\bibfnamefont
  {K.}~\bibnamefont {Izawa}}, \bibinfo {author} {\bibfnamefont {J.-P.}\
  \bibnamefont {Brison}},\ and\ \bibinfo {author} {\bibfnamefont
  {J.}~\bibnamefont {Flouquet}},\ }\href
  {https://doi.org/10.1103/PhysRevLett.115.046402} {\bibfield  {journal}
  {\bibinfo  {journal} {Phys. Rev. Lett.}\ }\textbf {\bibinfo {volume} {115}},\
  \bibinfo {pages} {046402} (\bibinfo {year} {2015})}\BibitemShut {NoStop}%
\bibitem [{\citenamefont {Ziman}(2001)}]{ziman2001electrons}%
  \BibitemOpen
  \bibfield  {author} {\bibinfo {author} {\bibfnamefont {J.~M.}\ \bibnamefont
  {Ziman}},\ }\href@noop {} {\emph {\bibinfo {title} {Electrons and phonons:
  the theory of transport phenomena in solids}}}\ (\bibinfo  {publisher}
  {Oxford university press},\ \bibinfo {year} {2001})\BibitemShut {NoStop}%
\bibitem [{\citenamefont {Mahan}(2013)}]{mahan2013many}%
  \BibitemOpen
  \bibfield  {author} {\bibinfo {author} {\bibfnamefont {G.~D.}\ \bibnamefont
  {Mahan}},\ }\href@noop {} {\emph {\bibinfo {title} {Many-particle physics}}}\
  (\bibinfo  {publisher} {Springer Science \& Business Media},\ \bibinfo {year}
  {2013})\BibitemShut {NoStop}%
\bibitem [{\citenamefont {Lussier}\ \emph {et~al.}(1994)\citenamefont
  {Lussier}, \citenamefont {Ellman},\ and\ \citenamefont
  {Taillefer}}]{PhysRevLett.73.3294}%
  \BibitemOpen
  \bibfield  {author} {\bibinfo {author} {\bibfnamefont {B.}~\bibnamefont
  {Lussier}}, \bibinfo {author} {\bibfnamefont {B.}~\bibnamefont {Ellman}},\
  and\ \bibinfo {author} {\bibfnamefont {L.}~\bibnamefont {Taillefer}},\ }\href
  {https://doi.org/10.1103/PhysRevLett.73.3294} {\bibfield  {journal} {\bibinfo
   {journal} {Phys. Rev. Lett.}\ }\textbf {\bibinfo {volume} {73}},\ \bibinfo
  {pages} {3294} (\bibinfo {year} {1994})}\BibitemShut {NoStop}%
\bibitem [{\citenamefont {Paglione}\ \emph {et~al.}(2005)\citenamefont
  {Paglione}, \citenamefont {Tanatar}, \citenamefont {Hawthorn}, \citenamefont
  {Hill}, \citenamefont {Ronning}, \citenamefont {Sutherland}, \citenamefont
  {Taillefer}, \citenamefont {Petrovic},\ and\ \citenamefont
  {Canfield}}]{PhysRevLett.94.216602}%
  \BibitemOpen
  \bibfield  {author} {\bibinfo {author} {\bibfnamefont {J.}~\bibnamefont
  {Paglione}}, \bibinfo {author} {\bibfnamefont {M.~A.}\ \bibnamefont
  {Tanatar}}, \bibinfo {author} {\bibfnamefont {D.~G.}\ \bibnamefont
  {Hawthorn}}, \bibinfo {author} {\bibfnamefont {R.~W.}\ \bibnamefont {Hill}},
  \bibinfo {author} {\bibfnamefont {F.}~\bibnamefont {Ronning}}, \bibinfo
  {author} {\bibfnamefont {M.}~\bibnamefont {Sutherland}}, \bibinfo {author}
  {\bibfnamefont {L.}~\bibnamefont {Taillefer}}, \bibinfo {author}
  {\bibfnamefont {C.}~\bibnamefont {Petrovic}},\ and\ \bibinfo {author}
  {\bibfnamefont {P.~C.}\ \bibnamefont {Canfield}},\ }\href
  {https://doi.org/10.1103/PhysRevLett.94.216602} {\bibfield  {journal}
  {\bibinfo  {journal} {Phys. Rev. Lett.}\ }\textbf {\bibinfo {volume} {94}},\
  \bibinfo {pages} {216602} (\bibinfo {year} {2005})}\BibitemShut {NoStop}%
\bibitem [{\citenamefont {Tanatar}\ \emph {et~al.}(2007)\citenamefont
  {Tanatar}, \citenamefont {Paglione}, \citenamefont {Petrovic},\ and\
  \citenamefont {Taillefer}}]{doi:10.1126/science.1140762}%
  \BibitemOpen
  \bibfield  {author} {\bibinfo {author} {\bibfnamefont {M.~A.}\ \bibnamefont
  {Tanatar}}, \bibinfo {author} {\bibfnamefont {J.}~\bibnamefont {Paglione}},
  \bibinfo {author} {\bibfnamefont {C.}~\bibnamefont {Petrovic}},\ and\
  \bibinfo {author} {\bibfnamefont {L.}~\bibnamefont {Taillefer}},\ }\href
  {https://doi.org/10.1126/science.1140762} {\bibfield  {journal} {\bibinfo
  {journal} {Science}\ }\textbf {\bibinfo {volume} {316}},\ \bibinfo {pages}
  {1320} (\bibinfo {year} {2007})}\BibitemShut {NoStop}%
\bibitem [{\citenamefont {Kim}\ and\ \citenamefont
  {P\'epin}(2009)}]{PhysRevLett.102.156404}%
  \BibitemOpen
  \bibfield  {author} {\bibinfo {author} {\bibfnamefont {K.-S.}\ \bibnamefont
  {Kim}}\ and\ \bibinfo {author} {\bibfnamefont {C.}~\bibnamefont {P\'epin}},\
  }\href {https://doi.org/10.1103/PhysRevLett.102.156404} {\bibfield  {journal}
  {\bibinfo  {journal} {Phys. Rev. Lett.}\ }\textbf {\bibinfo {volume} {102}},\
  \bibinfo {pages} {156404} (\bibinfo {year} {2009})}\BibitemShut {NoStop}%
\bibitem [{\citenamefont {Pfau}\ \emph {et~al.}(2012)\citenamefont {Pfau},
  \citenamefont {Hartmann}, \citenamefont {Stockert}, \citenamefont {Sun},
  \citenamefont {Lausberg}, \citenamefont {Brando}, \citenamefont {Friedemann},
  \citenamefont {Krellner}, \citenamefont {Geibel}, \citenamefont {Wirth},
  \citenamefont {Kirchner}, \citenamefont {Abrahams}, \citenamefont {Si},\ and\
  \citenamefont {Steglich}}]{Pfau2012}%
  \BibitemOpen
  \bibfield  {author} {\bibinfo {author} {\bibfnamefont {H.}~\bibnamefont
  {Pfau}}, \bibinfo {author} {\bibfnamefont {S.}~\bibnamefont {Hartmann}},
  \bibinfo {author} {\bibfnamefont {U.}~\bibnamefont {Stockert}}, \bibinfo
  {author} {\bibfnamefont {P.}~\bibnamefont {Sun}}, \bibinfo {author}
  {\bibfnamefont {S.}~\bibnamefont {Lausberg}}, \bibinfo {author}
  {\bibfnamefont {M.}~\bibnamefont {Brando}}, \bibinfo {author} {\bibfnamefont
  {S.}~\bibnamefont {Friedemann}}, \bibinfo {author} {\bibfnamefont
  {C.}~\bibnamefont {Krellner}}, \bibinfo {author} {\bibfnamefont
  {C.}~\bibnamefont {Geibel}}, \bibinfo {author} {\bibfnamefont
  {S.}~\bibnamefont {Wirth}}, \bibinfo {author} {\bibfnamefont
  {S.}~\bibnamefont {Kirchner}}, \bibinfo {author} {\bibfnamefont
  {E.}~\bibnamefont {Abrahams}}, \bibinfo {author} {\bibfnamefont
  {Q.}~\bibnamefont {Si}},\ and\ \bibinfo {author} {\bibfnamefont
  {F.}~\bibnamefont {Steglich}},\ }\href {https://doi.org/10.1038/nature11072}
  {\bibfield  {journal} {\bibinfo  {journal} {Nature}\ }\textbf {\bibinfo
  {volume} {484}},\ \bibinfo {pages} {493} (\bibinfo {year}
  {2012})}\BibitemShut {NoStop}%
\bibitem [{\citenamefont {Steglich}(2024)}]{cpl_41_12_127401}%
  \BibitemOpen
  \bibfield  {author} {\bibinfo {author} {\bibfnamefont {F.}~\bibnamefont
  {Steglich}},\ }\href {https://doi.org/10.1088/0256-307X/41/12/127401}
  {\bibfield  {journal} {\bibinfo  {journal} {Chin. Phys. Lett.}\ }\textbf
  {\bibinfo {volume} {41}},\ \bibinfo {pages} {127401} (\bibinfo {year}
  {2024})}\BibitemShut {NoStop}%
\bibitem [{\citenamefont {Kountz}\ \emph {et~al.}(2021)\citenamefont {Kountz},
  \citenamefont {Zhang}, \citenamefont {Straquadine}, \citenamefont {Singh},
  \citenamefont {Bachmann}, \citenamefont {Fisher}, \citenamefont {Kivelson},\
  and\ \citenamefont {Kapitulnik}}]{PhysRevB.104.L241109}%
  \BibitemOpen
  \bibfield  {author} {\bibinfo {author} {\bibfnamefont {E.~D.}\ \bibnamefont
  {Kountz}}, \bibinfo {author} {\bibfnamefont {J.}~\bibnamefont {Zhang}},
  \bibinfo {author} {\bibfnamefont {J.~A.~W.}\ \bibnamefont {Straquadine}},
  \bibinfo {author} {\bibfnamefont {A.~G.}\ \bibnamefont {Singh}}, \bibinfo
  {author} {\bibfnamefont {M.~D.}\ \bibnamefont {Bachmann}}, \bibinfo {author}
  {\bibfnamefont {I.~R.}\ \bibnamefont {Fisher}}, \bibinfo {author}
  {\bibfnamefont {S.~A.}\ \bibnamefont {Kivelson}},\ and\ \bibinfo {author}
  {\bibfnamefont {A.}~\bibnamefont {Kapitulnik}},\ }\href
  {https://doi.org/10.1103/PhysRevB.104.L241109} {\bibfield  {journal}
  {\bibinfo  {journal} {Phys. Rev. B}\ }\textbf {\bibinfo {volume} {104}},\
  \bibinfo {pages} {L241109} (\bibinfo {year} {2021})}\BibitemShut {NoStop}%
\bibitem [{\citenamefont {Gooth}\ \emph {et~al.}(2018)\citenamefont {Gooth},
  \citenamefont {Menges}, \citenamefont {Kumar}, \citenamefont {S{\"u}$\beta$},
  \citenamefont {Shekhar}, \citenamefont {Sun}, \citenamefont {Drechsler},
  \citenamefont {Zierold}, \citenamefont {Felser},\ and\ \citenamefont
  {Gotsmann}}]{Gooth2018}%
  \BibitemOpen
  \bibfield  {author} {\bibinfo {author} {\bibfnamefont {J.}~\bibnamefont
  {Gooth}}, \bibinfo {author} {\bibfnamefont {F.}~\bibnamefont {Menges}},
  \bibinfo {author} {\bibfnamefont {N.}~\bibnamefont {Kumar}}, \bibinfo
  {author} {\bibfnamefont {V.}~\bibnamefont {S{\"u}$\beta$}}, \bibinfo {author}
  {\bibfnamefont {C.}~\bibnamefont {Shekhar}}, \bibinfo {author} {\bibfnamefont
  {Y.}~\bibnamefont {Sun}}, \bibinfo {author} {\bibfnamefont {U.}~\bibnamefont
  {Drechsler}}, \bibinfo {author} {\bibfnamefont {R.}~\bibnamefont {Zierold}},
  \bibinfo {author} {\bibfnamefont {C.}~\bibnamefont {Felser}},\ and\ \bibinfo
  {author} {\bibfnamefont {B.}~\bibnamefont {Gotsmann}},\ }\href
  {https://doi.org/10.1038/s41467-018-06688-y} {\bibfield  {journal} {\bibinfo
  {journal} {Nature Communications}\ }\textbf {\bibinfo {volume} {9}},\
  \bibinfo {pages} {4093} (\bibinfo {year} {2018})}\BibitemShut {NoStop}%
\bibitem [{\citenamefont {Jaoui}\ \emph {et~al.}(2018)\citenamefont {Jaoui},
  \citenamefont {Fauqu{\'e}}, \citenamefont {Rischau}, \citenamefont {Subedi},
  \citenamefont {Fu}, \citenamefont {Gooth}, \citenamefont {Kumar},
  \citenamefont {S{\"u}{\ss}}, \citenamefont {Maslov}, \citenamefont {Felser},\
  and\ \citenamefont {Behnia}}]{Jaoui2018}%
  \BibitemOpen
  \bibfield  {author} {\bibinfo {author} {\bibfnamefont {A.}~\bibnamefont
  {Jaoui}}, \bibinfo {author} {\bibfnamefont {B.}~\bibnamefont {Fauqu{\'e}}},
  \bibinfo {author} {\bibfnamefont {C.~W.}\ \bibnamefont {Rischau}}, \bibinfo
  {author} {\bibfnamefont {A.}~\bibnamefont {Subedi}}, \bibinfo {author}
  {\bibfnamefont {C.}~\bibnamefont {Fu}}, \bibinfo {author} {\bibfnamefont
  {J.}~\bibnamefont {Gooth}}, \bibinfo {author} {\bibfnamefont
  {N.}~\bibnamefont {Kumar}}, \bibinfo {author} {\bibfnamefont
  {V.}~\bibnamefont {S{\"u}{\ss}}}, \bibinfo {author} {\bibfnamefont {D.~L.}\
  \bibnamefont {Maslov}}, \bibinfo {author} {\bibfnamefont {C.}~\bibnamefont
  {Felser}},\ and\ \bibinfo {author} {\bibfnamefont {K.}~\bibnamefont
  {Behnia}},\ }\href {https://doi.org/10.1038/s41535-018-0136-x} {\bibfield
  {journal} {\bibinfo  {journal} {npj Quantum Materials}\ }\textbf {\bibinfo
  {volume} {3}},\ \bibinfo {pages} {64} (\bibinfo {year} {2018})}\BibitemShut
  {NoStop}%
\bibitem [{\citenamefont {Wang}\ \emph {et~al.}(2025)\citenamefont {Wang},
  \citenamefont {Rao}, \citenamefont {Zhou}, \citenamefont {Zhu}, \citenamefont
  {Zhao}, \citenamefont {Chen}, \citenamefont {Li}, \citenamefont {Liang},
  \citenamefont {Xia},\ and\ \citenamefont {Sun}}]{Wang2025}%
  \BibitemOpen
  \bibfield  {author} {\bibinfo {author} {\bibfnamefont {Y.-Y.}\ \bibnamefont
  {Wang}}, \bibinfo {author} {\bibfnamefont {X.}~\bibnamefont {Rao}}, \bibinfo
  {author} {\bibfnamefont {Y.}~\bibnamefont {Zhou}}, \bibinfo {author}
  {\bibfnamefont {X.-D.}\ \bibnamefont {Zhu}}, \bibinfo {author} {\bibfnamefont
  {X.}~\bibnamefont {Zhao}}, \bibinfo {author} {\bibfnamefont {G.}~\bibnamefont
  {Chen}}, \bibinfo {author} {\bibfnamefont {N.}~\bibnamefont {Li}}, \bibinfo
  {author} {\bibfnamefont {H.}~\bibnamefont {Liang}}, \bibinfo {author}
  {\bibfnamefont {T.-L.}\ \bibnamefont {Xia}},\ and\ \bibinfo {author}
  {\bibfnamefont {X.-F.}\ \bibnamefont {Sun}},\ }\href
  {https://doi.org/10.1038/s41467-024-55141-w} {\bibfield  {journal} {\bibinfo
  {journal} {Nature Communications}\ }\textbf {\bibinfo {volume} {16}},\
  \bibinfo {pages} {53} (\bibinfo {year} {2025})}\BibitemShut {NoStop}%
\bibitem [{\citenamefont {Zhong}\ \emph {et~al.}(2025)\citenamefont {Zhong},
  \citenamefont {Jin}, \citenamefont {He}, \citenamefont {Wang}, \citenamefont
  {Zhou}, \citenamefont {Deng},\ and\ \citenamefont {Yang}}]{zhong2025phonon}%
  \BibitemOpen
  \bibfield  {author} {\bibinfo {author} {\bibfnamefont {L.}~\bibnamefont
  {Zhong}}, \bibinfo {author} {\bibfnamefont {X.}~\bibnamefont {Jin}}, \bibinfo
  {author} {\bibfnamefont {M.}~\bibnamefont {He}}, \bibinfo {author}
  {\bibfnamefont {R.}~\bibnamefont {Wang}}, \bibinfo {author} {\bibfnamefont
  {X.}~\bibnamefont {Zhou}}, \bibinfo {author} {\bibfnamefont {T.}~\bibnamefont
  {Deng}},\ and\ \bibinfo {author} {\bibfnamefont {X.}~\bibnamefont {Yang}},\
  }\href@noop {} {\bibfield  {journal} {\bibinfo  {journal} {arXiv preprint
  arXiv:2511.06290}\ } (\bibinfo {year} {2025})}\BibitemShut {NoStop}%
\bibitem [{\citenamefont {Principi}\ and\ \citenamefont
  {Vignale}(2015)}]{PhysRevLett.115.056603}%
  \BibitemOpen
  \bibfield  {author} {\bibinfo {author} {\bibfnamefont {A.}~\bibnamefont
  {Principi}}\ and\ \bibinfo {author} {\bibfnamefont {G.}~\bibnamefont
  {Vignale}},\ }\href {https://doi.org/10.1103/PhysRevLett.115.056603}
  {\bibfield  {journal} {\bibinfo  {journal} {Phys. Rev. Lett.}\ }\textbf
  {\bibinfo {volume} {115}},\ \bibinfo {pages} {056603} (\bibinfo {year}
  {2015})}\BibitemShut {NoStop}%
\bibitem [{\citenamefont {Crossno}\ \emph {et~al.}(2016)\citenamefont
  {Crossno}, \citenamefont {Shi}, \citenamefont {Wang}, \citenamefont {Liu},
  \citenamefont {Harzheim}, \citenamefont {Lucas}, \citenamefont {Sachdev},
  \citenamefont {Kim}, \citenamefont {Taniguchi}, \citenamefont {Watanabe},
  \citenamefont {Ohki},\ and\ \citenamefont
  {Fong}}]{doi:10.1126/science.aad0343}%
  \BibitemOpen
  \bibfield  {author} {\bibinfo {author} {\bibfnamefont {J.}~\bibnamefont
  {Crossno}}, \bibinfo {author} {\bibfnamefont {J.~K.}\ \bibnamefont {Shi}},
  \bibinfo {author} {\bibfnamefont {K.}~\bibnamefont {Wang}}, \bibinfo {author}
  {\bibfnamefont {X.}~\bibnamefont {Liu}}, \bibinfo {author} {\bibfnamefont
  {A.}~\bibnamefont {Harzheim}}, \bibinfo {author} {\bibfnamefont
  {A.}~\bibnamefont {Lucas}}, \bibinfo {author} {\bibfnamefont
  {S.}~\bibnamefont {Sachdev}}, \bibinfo {author} {\bibfnamefont
  {P.}~\bibnamefont {Kim}}, \bibinfo {author} {\bibfnamefont {T.}~\bibnamefont
  {Taniguchi}}, \bibinfo {author} {\bibfnamefont {K.}~\bibnamefont {Watanabe}},
  \bibinfo {author} {\bibfnamefont {T.~A.}\ \bibnamefont {Ohki}},\ and\
  \bibinfo {author} {\bibfnamefont {K.~C.}\ \bibnamefont {Fong}},\ }\href
  {https://doi.org/10.1126/science.aad0343} {\bibfield  {journal} {\bibinfo
  {journal} {Science}\ }\textbf {\bibinfo {volume} {351}},\ \bibinfo {pages}
  {1058} (\bibinfo {year} {2016})}\BibitemShut {NoStop}%
\bibitem [{\citenamefont {Mahajan}\ \emph {et~al.}(2013)\citenamefont
  {Mahajan}, \citenamefont {Barkeshli},\ and\ \citenamefont
  {Hartnoll}}]{PhysRevB.88.125107}%
  \BibitemOpen
  \bibfield  {author} {\bibinfo {author} {\bibfnamefont {R.}~\bibnamefont
  {Mahajan}}, \bibinfo {author} {\bibfnamefont {M.}~\bibnamefont {Barkeshli}},\
  and\ \bibinfo {author} {\bibfnamefont {S.~A.}\ \bibnamefont {Hartnoll}},\
  }\href {https://doi.org/10.1103/PhysRevB.88.125107} {\bibfield  {journal}
  {\bibinfo  {journal} {Phys. Rev. B}\ }\textbf {\bibinfo {volume} {88}},\
  \bibinfo {pages} {125107} (\bibinfo {year} {2013})}\BibitemShut {NoStop}%
\bibitem [{\citenamefont {Lucas}\ and\ \citenamefont
  {Das~Sarma}(2018)}]{PhysRevB.97.245128}%
  \BibitemOpen
  \bibfield  {author} {\bibinfo {author} {\bibfnamefont {A.}~\bibnamefont
  {Lucas}}\ and\ \bibinfo {author} {\bibfnamefont {S.}~\bibnamefont
  {Das~Sarma}},\ }\href {https://doi.org/10.1103/PhysRevB.97.245128} {\bibfield
   {journal} {\bibinfo  {journal} {Phys. Rev. B}\ }\textbf {\bibinfo {volume}
  {97}},\ \bibinfo {pages} {245128} (\bibinfo {year} {2018})}\BibitemShut
  {NoStop}%
\bibitem [{\citenamefont {Coulter}\ \emph {et~al.}(2018)\citenamefont
  {Coulter}, \citenamefont {Sundararaman},\ and\ \citenamefont
  {Narang}}]{PhysRevB.98.115130}%
  \BibitemOpen
  \bibfield  {author} {\bibinfo {author} {\bibfnamefont {J.}~\bibnamefont
  {Coulter}}, \bibinfo {author} {\bibfnamefont {R.}~\bibnamefont
  {Sundararaman}},\ and\ \bibinfo {author} {\bibfnamefont {P.}~\bibnamefont
  {Narang}},\ }\href {https://doi.org/10.1103/PhysRevB.98.115130} {\bibfield
  {journal} {\bibinfo  {journal} {Phys. Rev. B}\ }\textbf {\bibinfo {volume}
  {98}},\ \bibinfo {pages} {115130} (\bibinfo {year} {2018})}\BibitemShut
  {NoStop}%
\bibitem [{\citenamefont {Zarenia}\ \emph {et~al.}(2019)\citenamefont
  {Zarenia}, \citenamefont {Principi},\ and\ \citenamefont
  {Vignale}}]{Zarenia_2019}%
  \BibitemOpen
  \bibfield  {author} {\bibinfo {author} {\bibfnamefont {M.}~\bibnamefont
  {Zarenia}}, \bibinfo {author} {\bibfnamefont {A.}~\bibnamefont {Principi}},\
  and\ \bibinfo {author} {\bibfnamefont {G.}~\bibnamefont {Vignale}},\ }\href
  {https://doi.org/10.1088/2053-1583/ab1ad9} {\bibfield  {journal} {\bibinfo
  {journal} {2D Materials}\ }\textbf {\bibinfo {volume} {6}},\ \bibinfo {pages}
  {035024} (\bibinfo {year} {2019})}\BibitemShut {NoStop}%
\bibitem [{\citenamefont {Jaoui}\ \emph {et~al.}(2021)\citenamefont {Jaoui},
  \citenamefont {Fauqu{\'e}},\ and\ \citenamefont {Behnia}}]{Jaoui2021}%
  \BibitemOpen
  \bibfield  {author} {\bibinfo {author} {\bibfnamefont {A.}~\bibnamefont
  {Jaoui}}, \bibinfo {author} {\bibfnamefont {B.}~\bibnamefont {Fauqu{\'e}}},\
  and\ \bibinfo {author} {\bibfnamefont {K.}~\bibnamefont {Behnia}},\ }\href
  {https://doi.org/10.1038/s41467-020-20420-9} {\bibfield  {journal} {\bibinfo
  {journal} {Nature Communications}\ }\textbf {\bibinfo {volume} {12}},\
  \bibinfo {pages} {195} (\bibinfo {year} {2021})}\BibitemShut {NoStop}%
\bibitem [{\citenamefont {Fritz}\ and\ \citenamefont
  {Scaffidi}(2024)}]{annurev}%
  \BibitemOpen
  \bibfield  {author} {\bibinfo {author} {\bibfnamefont {L.}~\bibnamefont
  {Fritz}}\ and\ \bibinfo {author} {\bibfnamefont {T.}~\bibnamefont
  {Scaffidi}},\ }\href
  {https://doi.org/https://doi.org/10.1146/annurev-conmatphys-040521-042014}
  {\bibfield  {journal} {\bibinfo  {journal} {Annual Review of Condensed Matter
  Physics}\ }\textbf {\bibinfo {volume} {15}},\ \bibinfo {pages} {17} (\bibinfo
  {year} {2024})}\BibitemShut {NoStop}%
\bibitem [{\citenamefont {Hui}\ and\ \citenamefont {Skinner}(2025)}]{Hui_2025}%
  \BibitemOpen
  \bibfield  {author} {\bibinfo {author} {\bibfnamefont {A.}~\bibnamefont
  {Hui}}\ and\ \bibinfo {author} {\bibfnamefont {B.}~\bibnamefont {Skinner}},\
  }\href {https://doi.org/10.1088/1361-648X/adfbcd} {\bibfield  {journal}
  {\bibinfo  {journal} {Journal of Physics: Condensed Matter}\ }\textbf
  {\bibinfo {volume} {37}},\ \bibinfo {pages} {363001} (\bibinfo {year}
  {2025})}\BibitemShut {NoStop}%
\bibitem [{SM()}]{SM}%
  \BibitemOpen
  \href@noop {} {}\bibinfo {note} {See Supplemental Material at
  ...}\BibitemShut {Stop}%
\bibitem [{\citenamefont {Qiao}\ \emph {et~al.}(2010)\citenamefont {Qiao},
  \citenamefont {Yang}, \citenamefont {Feng}, \citenamefont {Tse},
  \citenamefont {Ding}, \citenamefont {Yao}, \citenamefont {Wang},\ and\
  \citenamefont {Niu}}]{PhysRevB.82.161414}%
  \BibitemOpen
  \bibfield  {author} {\bibinfo {author} {\bibfnamefont {Z.}~\bibnamefont
  {Qiao}}, \bibinfo {author} {\bibfnamefont {S.~A.}\ \bibnamefont {Yang}},
  \bibinfo {author} {\bibfnamefont {W.}~\bibnamefont {Feng}}, \bibinfo {author}
  {\bibfnamefont {W.-K.}\ \bibnamefont {Tse}}, \bibinfo {author} {\bibfnamefont
  {J.}~\bibnamefont {Ding}}, \bibinfo {author} {\bibfnamefont {Y.}~\bibnamefont
  {Yao}}, \bibinfo {author} {\bibfnamefont {J.}~\bibnamefont {Wang}},\ and\
  \bibinfo {author} {\bibfnamefont {Q.}~\bibnamefont {Niu}},\ }\href
  {https://doi.org/10.1103/PhysRevB.82.161414} {\bibfield  {journal} {\bibinfo
  {journal} {Phys. Rev. B}\ }\textbf {\bibinfo {volume} {82}},\ \bibinfo
  {pages} {161414} (\bibinfo {year} {2010})}\BibitemShut {NoStop}%
\bibitem [{\citenamefont {Rachel}\ and\ \citenamefont
  {Le~Hur}(2010)}]{PhysRevB.82.075106}%
  \BibitemOpen
  \bibfield  {author} {\bibinfo {author} {\bibfnamefont {S.}~\bibnamefont
  {Rachel}}\ and\ \bibinfo {author} {\bibfnamefont {K.}~\bibnamefont
  {Le~Hur}},\ }\href {https://doi.org/10.1103/PhysRevB.82.075106} {\bibfield
  {journal} {\bibinfo  {journal} {Phys. Rev. B}\ }\textbf {\bibinfo {volume}
  {82}},\ \bibinfo {pages} {075106} (\bibinfo {year} {2010})}\BibitemShut
  {NoStop}%
\bibitem [{\citenamefont {Tse}\ \emph {et~al.}(2011)\citenamefont {Tse},
  \citenamefont {Qiao}, \citenamefont {Yao}, \citenamefont {MacDonald},\ and\
  \citenamefont {Niu}}]{PhysRevB.83.155447}%
  \BibitemOpen
  \bibfield  {author} {\bibinfo {author} {\bibfnamefont {W.-K.}\ \bibnamefont
  {Tse}}, \bibinfo {author} {\bibfnamefont {Z.}~\bibnamefont {Qiao}}, \bibinfo
  {author} {\bibfnamefont {Y.}~\bibnamefont {Yao}}, \bibinfo {author}
  {\bibfnamefont {A.~H.}\ \bibnamefont {MacDonald}},\ and\ \bibinfo {author}
  {\bibfnamefont {Q.}~\bibnamefont {Niu}},\ }\href
  {https://doi.org/10.1103/PhysRevB.83.155447} {\bibfield  {journal} {\bibinfo
  {journal} {Phys. Rev. B}\ }\textbf {\bibinfo {volume} {83}},\ \bibinfo
  {pages} {155447} (\bibinfo {year} {2011})}\BibitemShut {NoStop}%
\bibitem [{\citenamefont {Sorella}\ and\ \citenamefont
  {Tosatti}(1992)}]{S_Sorella_1992}%
  \BibitemOpen
  \bibfield  {author} {\bibinfo {author} {\bibfnamefont {S.}~\bibnamefont
  {Sorella}}\ and\ \bibinfo {author} {\bibfnamefont {E.}~\bibnamefont
  {Tosatti}},\ }\href {https://doi.org/10.1209/0295-5075/19/8/007} {\bibfield
  {journal} {\bibinfo  {journal} {Europhysics Letters}\ }\textbf {\bibinfo
  {volume} {19}},\ \bibinfo {pages} {699} (\bibinfo {year} {1992})}\BibitemShut
  {NoStop}%
\bibitem [{\citenamefont {Jafari}(2009)}]{Jafari2009}%
  \BibitemOpen
  \bibfield  {author} {\bibinfo {author} {\bibfnamefont {S.~A.}\ \bibnamefont
  {Jafari}},\ }\href {https://doi.org/10.1140/epjb/e2009-00128-1} {\bibfield
  {journal} {\bibinfo  {journal} {The European Physical Journal B}\ }\textbf
  {\bibinfo {volume} {68}},\ \bibinfo {pages} {537} (\bibinfo {year}
  {2009})}\BibitemShut {NoStop}%
\bibitem [{\citenamefont {Xu}\ \emph {et~al.}(2020)\citenamefont {Xu},
  \citenamefont {Li}, \citenamefont {Lu}, \citenamefont {Collignon},
  \citenamefont {Fu}, \citenamefont {Koo}, \citenamefont {Fauqué},
  \citenamefont {Yan}, \citenamefont {Zhu},\ and\ \citenamefont
  {Behnia}}]{doi:10.1126/sciadv.aaz3522}%
  \BibitemOpen
  \bibfield  {author} {\bibinfo {author} {\bibfnamefont {L.}~\bibnamefont
  {Xu}}, \bibinfo {author} {\bibfnamefont {X.}~\bibnamefont {Li}}, \bibinfo
  {author} {\bibfnamefont {X.}~\bibnamefont {Lu}}, \bibinfo {author}
  {\bibfnamefont {C.}~\bibnamefont {Collignon}}, \bibinfo {author}
  {\bibfnamefont {H.}~\bibnamefont {Fu}}, \bibinfo {author} {\bibfnamefont
  {J.}~\bibnamefont {Koo}}, \bibinfo {author} {\bibfnamefont {B.}~\bibnamefont
  {Fauqué}}, \bibinfo {author} {\bibfnamefont {B.}~\bibnamefont {Yan}},
  \bibinfo {author} {\bibfnamefont {Z.}~\bibnamefont {Zhu}},\ and\ \bibinfo
  {author} {\bibfnamefont {K.}~\bibnamefont {Behnia}},\ }\href
  {https://doi.org/10.1126/sciadv.aaz3522} {\bibfield  {journal} {\bibinfo
  {journal} {Science Advances}\ }\textbf {\bibinfo {volume} {6}},\ \bibinfo
  {pages} {eaaz3522} (\bibinfo {year} {2020})}\BibitemShut {NoStop}%
\bibitem [{\citenamefont {Ding}\ \emph {et~al.}(2021)\citenamefont {Ding},
  \citenamefont {Koo}, \citenamefont {Yi}, \citenamefont {Xu}, \citenamefont
  {Zuo}, \citenamefont {Yang}, \citenamefont {Shi}, \citenamefont {Yan},
  \citenamefont {Behnia},\ and\ \citenamefont {Zhu}}]{Ding_2021}%
  \BibitemOpen
  \bibfield  {author} {\bibinfo {author} {\bibfnamefont {L.}~\bibnamefont
  {Ding}}, \bibinfo {author} {\bibfnamefont {J.}~\bibnamefont {Koo}}, \bibinfo
  {author} {\bibfnamefont {C.}~\bibnamefont {Yi}}, \bibinfo {author}
  {\bibfnamefont {L.}~\bibnamefont {Xu}}, \bibinfo {author} {\bibfnamefont
  {H.}~\bibnamefont {Zuo}}, \bibinfo {author} {\bibfnamefont {M.}~\bibnamefont
  {Yang}}, \bibinfo {author} {\bibfnamefont {Y.}~\bibnamefont {Shi}}, \bibinfo
  {author} {\bibfnamefont {B.}~\bibnamefont {Yan}}, \bibinfo {author}
  {\bibfnamefont {K.}~\bibnamefont {Behnia}},\ and\ \bibinfo {author}
  {\bibfnamefont {Z.}~\bibnamefont {Zhu}},\ }\href
  {https://doi.org/10.1088/1361-6463/ac1c2b} {\bibfield  {journal} {\bibinfo
  {journal} {Journal of Physics D: Applied Physics}\ }\textbf {\bibinfo
  {volume} {54}},\ \bibinfo {pages} {454003} (\bibinfo {year}
  {2021})}\BibitemShut {NoStop}%
\bibitem [{\citenamefont {Zhou}\ \emph {et~al.}(2022)\citenamefont {Zhou},
  \citenamefont {Liu}, \citenamefont {Wu}, \citenamefont {Jiang}, \citenamefont
  {Shi}, \citenamefont {Li}, \citenamefont {Sui}, \citenamefont {Hu},\ and\
  \citenamefont {Luo}}]{PhysRevB.105.205104}%
  \BibitemOpen
  \bibfield  {author} {\bibinfo {author} {\bibfnamefont {X.}~\bibnamefont
  {Zhou}}, \bibinfo {author} {\bibfnamefont {H.}~\bibnamefont {Liu}}, \bibinfo
  {author} {\bibfnamefont {W.}~\bibnamefont {Wu}}, \bibinfo {author}
  {\bibfnamefont {K.}~\bibnamefont {Jiang}}, \bibinfo {author} {\bibfnamefont
  {Y.}~\bibnamefont {Shi}}, \bibinfo {author} {\bibfnamefont {Z.}~\bibnamefont
  {Li}}, \bibinfo {author} {\bibfnamefont {Y.}~\bibnamefont {Sui}}, \bibinfo
  {author} {\bibfnamefont {J.}~\bibnamefont {Hu}},\ and\ \bibinfo {author}
  {\bibfnamefont {J.}~\bibnamefont {Luo}},\ }\href
  {https://doi.org/10.1103/PhysRevB.105.205104} {\bibfield  {journal} {\bibinfo
   {journal} {Phys. Rev. B}\ }\textbf {\bibinfo {volume} {105}},\ \bibinfo
  {pages} {205104} (\bibinfo {year} {2022})}\BibitemShut {NoStop}%
\bibitem [{\citenamefont {Zhang}\ \emph {et~al.}(2024)\citenamefont {Zhang},
  \citenamefont {Chen}, \citenamefont {Zheng}, \citenamefont {Yu},
  \citenamefont {Shi}, \citenamefont {Zhu}, \citenamefont {Chan}, \citenamefont
  {Jenkins}, \citenamefont {Ying}, \citenamefont {Xiang}, \citenamefont
  {Chen},\ and\ \citenamefont {Li}}]{Zhang2024}%
  \BibitemOpen
  \bibfield  {author} {\bibinfo {author} {\bibfnamefont {D.}~\bibnamefont
  {Zhang}}, \bibinfo {author} {\bibfnamefont {K.-W.}\ \bibnamefont {Chen}},
  \bibinfo {author} {\bibfnamefont {G.}~\bibnamefont {Zheng}}, \bibinfo
  {author} {\bibfnamefont {F.}~\bibnamefont {Yu}}, \bibinfo {author}
  {\bibfnamefont {M.}~\bibnamefont {Shi}}, \bibinfo {author} {\bibfnamefont
  {Y.}~\bibnamefont {Zhu}}, \bibinfo {author} {\bibfnamefont {A.}~\bibnamefont
  {Chan}}, \bibinfo {author} {\bibfnamefont {K.}~\bibnamefont {Jenkins}},
  \bibinfo {author} {\bibfnamefont {J.}~\bibnamefont {Ying}}, \bibinfo {author}
  {\bibfnamefont {Z.}~\bibnamefont {Xiang}}, \bibinfo {author} {\bibfnamefont
  {X.}~\bibnamefont {Chen}},\ and\ \bibinfo {author} {\bibfnamefont
  {L.}~\bibnamefont {Li}},\ }\href {https://doi.org/10.1038/s41467-024-50336-7}
  {\bibfield  {journal} {\bibinfo  {journal} {Nature Communications}\ }\textbf
  {\bibinfo {volume} {15}},\ \bibinfo {pages} {6224} (\bibinfo {year}
  {2024})}\BibitemShut {NoStop}%
\bibitem [{\citenamefont {Matsumoto}\ and\ \citenamefont
  {Murakami}(2011)}]{PhysRevLett.106.197202}%
  \BibitemOpen
  \bibfield  {author} {\bibinfo {author} {\bibfnamefont {R.}~\bibnamefont
  {Matsumoto}}\ and\ \bibinfo {author} {\bibfnamefont {S.}~\bibnamefont
  {Murakami}},\ }\href {https://doi.org/10.1103/PhysRevLett.106.197202}
  {\bibfield  {journal} {\bibinfo  {journal} {Phys. Rev. Lett.}\ }\textbf
  {\bibinfo {volume} {106}},\ \bibinfo {pages} {197202} (\bibinfo {year}
  {2011})}\BibitemShut {NoStop}%
\bibitem [{\citenamefont {Zhang}(2016)}]{Zhang_2016}%
  \BibitemOpen
  \bibfield  {author} {\bibinfo {author} {\bibfnamefont {L.}~\bibnamefont
  {Zhang}},\ }\href {https://doi.org/10.1088/1367-2630/18/10/103039} {\bibfield
   {journal} {\bibinfo  {journal} {New Journal of Physics}\ }\textbf {\bibinfo
  {volume} {18}},\ \bibinfo {pages} {103039} (\bibinfo {year}
  {2016})}\BibitemShut {NoStop}%
\end{thebibliography}
%

\clearpage
\onecolumngrid

\setcounter{page}{1}
\setcounter{figure}{0}
\setcounter{equation}{0}
\setcounter{section}{0}

\renewcommand{\thepage}{S\arabic{page}}
\renewcommand{\thesection}{S\arabic{section}}
\renewcommand{\thetable}{S\arabic{table}}
\renewcommand{\thefigure}{S\arabic{figure}}
\renewcommand{\theequation}{S\arabic{equation}}

\begin{center}
    \textbf{\large Supplemental Material for: \\ Intrinsic violation of the Wiedemann-Franz law in interacting systems}
    \\[1em]
    YuanDong Wang\textsuperscript{1,*} and Zhen-Gang Zhu\textsuperscript{2,$\dagger$} \\
    \vspace{0.5em}
    \textit{\textsuperscript{1}Department of Applied Physics, College of Science, China Agricultural University, Qinghua East Road, Beijing 100083, China}\\
    \textit{\textsuperscript{2}School of Electronic, Electrical and Communication Engineering, University of Chinese Academy of Sciences, Beijing 100049, China}
\end{center}

\section{Derivation of the Modified Wiedemann-Franz Ratio}\label{appa}

Here we provide the  derivation of the thermal and electrical conductivity in the presence of a temperature-dependent band structure, and explicitly derive the modified Wiedemann-Franz ratio. We start from the  Boltzman equation with the relaxation time approximation  for the distribution of electrons in absence of electric field:
\begin{equation}\label{boltz}
\bm{v}\cdot \bm{\nabla}_{\bm{r}}f_0 +\frac{e\bm{E}}{\hbar}\cdot \bm{\nabla}_{\bm{k}}f_0 = -\frac{\delta f}{\tau}.
\end{equation}
where $\tau$ is the relaxation time, $v$ is the group velocity.
And the semiclassical equations of motion
\begin{equation}\label{dot-rk}
\begin{aligned}
\dot{\bm{r}} =& \bm{v}_{\bm{k}}+\frac{e}{\hbar}\bm{E}\times \bm{\Omega}_{\bm{k}},\\
\dot{\bm{k}} =& -\frac{e}{\hbar}\bm{E}.
\end{aligned}
\end{equation}
The transport electric current is
\begin{equation}\label{je}
    \bm{J}^{\text{tr}} = -e\int  \dot{\bm{r}}  f(\bm{r},\bm{k}) [d\bm{k}],
\end{equation}
where $[d\bm{k}]$ is shorthand for $d\bm{k}/(2\pi)^3$.
The transport thermal current  $\bm{J}^{\text{tr}}_Q$ is defined as the flow of heat energy \cite{PhysRevLett.106.197202,Zhang_2016}:
\begin{equation}\label{jq}
    \bm{J}^{\text{tr}}_Q = \int (\varepsilon_{\bm{k}} - \mu)  \dot{\bm{r}} f(\bm{r},\bm{k}) [d\bm{k}] - \bm{\nabla}\times \bm{M}^{\text{edge}}_Q,
\end{equation}
where $\bm{M}^{\text{edge}}_Q$ is the energy thermal magnetization at the edge.
In general, the  electrical conductivity tensor can be written as
\begin{equation}\label{elec-conduc}
    \hat{\sigma} = \int \hat{\Sigma}(\varepsilon) \left(-\frac{\partial f_0}{\partial \varepsilon}\right) d\varepsilon,
\end{equation}
where $f_0(\varepsilon) = \left[e^{(\varepsilon-\mu)/k_B T} + 1\right]^{-1}$ is the Fermi-Dirac distribution, and $\hat{\Sigma}$ is the transport kernel.

To evaluate transport integrals analytically at low temperatures, we employ the Sommerfeld expansion. For a generic smooth function $H(\varepsilon)$, the integral over the derivative of the Fermi function is approximated as:
\begin{equation}\label{eq:sommerfeld_gen}
    \int_{-\infty}^{\infty} H(\varepsilon) \left( -\frac{\partial f_0}{\partial \varepsilon} \right) d\varepsilon \approx H(\mu) + \frac{\pi^2}{6}(k_B T)^2 H''(\mu) + \dots
\end{equation}
Applying Eq.~\eqref{eq:sommerfeld_gen} to the electrical conductivity \Eq{elec-conduc}, where $H(\varepsilon) = \Sigma(\varepsilon)$, and retaining only the leading order term (as the second-order term is negligible for metallic $\sigma$), we obtain:
\begin{equation}\label{ec}
    \hat{\sigma} \approx \Sigma(\mu).
\end{equation}
In previous studies, the temperature dependency of the band structure on temperature is omitted. However, when interaction is present, this dependency can not be omitted, especially near the phase transition point that temperature strongly alternates dispersions, and we must reconsider the low temperature expansion.
 In the presence of interaction, the total temperature derivative is:
\begin{equation}\label{deriv}
    \frac{\partial f_0}{\partial T} = \frac{\partial f_0}{\partial \varepsilon} \left( -\frac{\varepsilon - \mu}{T} + \frac{\partial \varepsilon}{\partial T} \right) .
\end{equation}
Here, the term $\partial \varepsilon/ \partial T$ represents the interaction-induced energy drift (IED). Noting that $\frac{\partial f_0}{\partial T} = \frac{\partial f_0}{\partial \varepsilon} \frac{\varepsilon - \mu}{T}$ is valid only if $\varepsilon$ is $T$-independent. Consequently, the thermal conductivity tensor is given by (see Sec. \ref{suba} and \ref{subb})
\begin{equation}
    \hat{\kappa} = \int \frac{\hat{\Sigma}(\varepsilon)}{e^2}(\varepsilon - \mu) \left( \frac{\varepsilon - \mu}{T} - \frac{\partial \varepsilon}{\partial T} \right) \left( - \frac{\partial f_0}{\partial \varepsilon} \right) d\varepsilon.
\end{equation}
This integral can be separated into the standard Wiedemann-Franz term and an interaction correction term. We define two auxiliary functions:
\begin{align}
    H_1(\varepsilon) &= \frac{1}{e^2 T} \hat{\Sigma}(\varepsilon) (\varepsilon - \mu)^2, \\
    H_2(\varepsilon) &= - \frac{1}{e^2} \hat{\Sigma}(\varepsilon) (\varepsilon - \mu) \frac{\partial \varepsilon}{\partial T}.
\end{align}
The total thermal conductivity is then
\begin{equation}
\hat{\kappa} = \int [H_1(\varepsilon) + H_2(\varepsilon)] \left( - \frac{\partial f_0}{\partial \varepsilon} \right) d\varepsilon.
\end{equation}
 We evaluate these using the Sommerfeld expansion Eq.~\eqref{eq:sommerfeld_gen}.

{\it{Standard Term ($H_1$)}}.
For $H_1$, the zeroth-order term vanishes because $(\varepsilon-\mu)^2|_{\mu} = 0$. The leading contribution comes from the second derivative, we have:
\begin{equation}
\begin{aligned}
    \hat{\kappa}_{\text{WF}} = \frac{\pi^2}{6}(k_B T)^2 H_1''(\mu) = \frac{\pi^2}{3}(k_B T)^2 \frac{\hat{\Sigma}(\mu)}{e^2 T}.
\end{aligned}
\end{equation}
Making use of \Eq{ec}, we have
\begin{equation}
\begin{aligned}
    \hat{\kappa}_{\text{WF}}  = \frac{\pi^2 k_B^2 T}{3 e^2} \hat{\Sigma}(\mu) = L_0 T \hat{\sigma}.
\end{aligned}
\end{equation}
where the Lorentz number is defined as $L_0 ={\pi^2 k_B^2 }/{(3 e^2)}  $.
This recovers the standard WF law.

{\it{Interaction Term ($H_2$)}}.
For $H_2$, the zeroth-order term also vanishes. We must evaluate the second derivative $H_2''(\mu)$. After simple algebra, we have \footnote{Note that $H_2(\varepsilon)$ has the form $(\varepsilon-\mu) \cdot \hat{A}(\varepsilon)$, where $\hat{A}(\varepsilon) = -\frac{1}{e^2}\hat{\Sigma}(\varepsilon)\frac{\partial \varepsilon}{\partial T}$. The second derivative at $\varepsilon=\mu$ is simply $2\hat{A}'(\mu)$}
\begin{equation}
    \hat{\kappa}_{\text{Int}} = \frac{\pi^2}{6}(k_B T)^2 \cdot 2 \left[ \frac{d}{d\varepsilon} \left( - \frac{1}{e^2} \hat{\Sigma}(\varepsilon) \frac{\partial \varepsilon}{\partial T} \right) \right]_{\mu}.
\end{equation}
Expanding the derivative:
\begin{equation}
    \hat{\kappa}_{\text{Int}}= - \frac{\pi^2 k_B^2 T^2}{3 e^2} \left[ \frac{\partial \hat{\Sigma}}{\partial \varepsilon} \frac{\partial \varepsilon}{\partial T} + \hat{\Sigma} \frac{\partial^2 \varepsilon}{\partial \varepsilon \partial T} \right]_{\mu}.
\end{equation}
Under the rigid band shift approximation, we assume the band shape does not deform significantly with temperature, implying $\frac{\partial}{\partial \varepsilon} ( \frac{\partial \varepsilon}{\partial T} ) \approx 0$. The expression simplifies to:
\begin{equation}
    \hat{\kappa}_{\text{Int}} = - \frac{\pi^2 k_B^2 T^2}{3 e^2} \left( \frac{\partial \hat{\Sigma}}{\partial \varepsilon} \right)_{\mu} \left( \frac{\partial \varepsilon}{\partial T} \right)_{\mu}.
\end{equation}


 Combining the contributions $\hat{\kappa} = \hat{\kappa}_{\text{WF}} + \hat{\kappa}_{\text{Int}}$, the Lorenz ratio becomes:
\begin{equation}\label{vio}
    \frac{\hat{\kappa}}{\hat{\sigma} T} = \frac{L_0 T \hat{\sigma}  + \hat{\kappa}_{\text{Int}}}{\sigma T} = L_0 \left( 1 + \frac{\hat{\kappa}_{\text{Int}}}{L_0 T \hat{\sigma}} \right).
\end{equation}
Substituting $\sigma \approx \Sigma (\mu)$ and the expression for $\kappa_{\text{Int}}$:
\begin{equation}
\begin{aligned}
    \frac{\hat{\kappa}}{\hat{\sigma} T} =& L_0 \left[ 1 - \frac{1}{L_0 T \hat{\Sigma}(\mu)} \frac{\pi^2 k_B^2 T^2}{3 e^2}  \left(\frac{\partial \hat{\Sigma}}{\partial \varepsilon} \frac{\partial \varepsilon}{\partial T}\right)_{\mu} \right]\\
    =&L_0 \left[ 1 -  T \frac{1}{\hat{\Sigma}(\mu)} \left(\frac{\partial \hat{\Sigma}}{\partial \varepsilon} \frac{\partial \varepsilon}{\partial T}\right)_{\mu}\right].
\end{aligned}
\end{equation}
Finally, we relate this to the Mott formula for the  Mott ratio tensor,
\begin{equation}\label{sb}
\hat{\mathcal{S}} = \frac{\pi^2 k_B^2 T}{3 e}\frac{1}{\hat{\Sigma}(\mu)} \left(\frac{\partial \hat{\Sigma}}{\partial \varepsilon}\right)_\mu .
\end{equation}
Substituting \Eq{sb} into \Eq{vio} yields the final  form:
\begin{equation}\label{v-wf}
    \frac{\hat{\kappa}}{\hat{\sigma} T} = L_0 \hat{\gamma}(T)
\end{equation}
where the deviation coefficient $\gamma(T)$ is
\begin{equation}\label{devia-all}
\hat{\gamma}(T) =  1 - \frac{1}{L_0 e} \hat{\mathcal{S}} \left( \frac{\partial \varepsilon}{\partial T} \right)_{\mu} .
\end{equation}
This result demonstrates that a finite Mott ratio tensor, combined with a interaction-induced energy drift, leads to a deviation from the universal Lorenz number. This drift $\partial \varepsilon_{\bm{k}} / \partial T$ acts as an effective interaction-induced driving force that distinguishes heat transport from charge transport. Noting that in above derivations, we have not distinguish the response tensors with different directions. The longitudinal and transverse responses, owing to different transport kernels, leading to significant differences on the deviation of WF law. According to \Eq{devia-all}, this difference manifests itself in Mott ratio tensor elements. The general deviation formula derived in \Eq{devia-all} applies to any transport channel. However, the physical origin of the transport kernel $\Sigma(\epsilon)$ and thus the nature of the violation differs fundamentally between longitudinal and transverse responses. Here we derive the explicit forms for both cases.

\subsection{Longitudinal}\label{suba}

Applying \Eq{boltz} and \Eq{dot-rk} to \Eq{je}, it is straightforward to obtain  the transport kernel function for longitudinal transport (for instance, the $x$-direction)
\begin{equation}
    \Sigma_{xx}(\epsilon) = e^2 \tau(\epsilon) \int \frac{d^2k}{(2\pi)^2} v_x^2(\mathbf{k}) \delta(\epsilon - \epsilon_{\mathbf{k}}),
\end{equation}
which can be simplified as $\Sigma_{xx}(\epsilon) = e^2 \tau(\epsilon)  v_x^2(\varepsilon) N(\epsilon)$, where $N(\epsilon)$ is the density of states (DOS). In low temperature, there is $\sigma^{xx}\approx \Sigma^{xx}(\mu)$.

In the presence of temperature gradient,
\begin{equation}\label{d-f}
     \bm{\nabla}_{\bm{r}} f_0 = \frac{\partial f_0}{\partial \varepsilon} \left( -\frac{\varepsilon - \mu}{T} + \frac{\partial \varepsilon}{\partial T} \right) \bm{\nabla} T.
\end{equation}
Combining \Eq{d-f}, \Eq{dot-rk} and \Eq{jq}, and assuming the system is Fermi liquid, where the relaxation times for charge and heat are not separated, i.e., $\tau_{\text{charge}} = \tau_{\text{heat}} = \tau$).
We obtain
\begin{equation}
    \kappa^{xx} = \int \frac{\Sigma^{xx}(\varepsilon)}{e^2}(\varepsilon - \mu) \left( \frac{\varepsilon - \mu}{T} - \frac{\partial \varepsilon}{\partial T} \right) \left( - \frac{\partial f_0}{\partial \varepsilon} \right) d\varepsilon.
\end{equation}
Substituting this into the deviation factor $\gamma_{xx}$:
\begin{equation}\label{devia-longi}
{\gamma}_{xx} =  1 - \frac{1}{L_0 e} \mathcal{S}_{xx} \left( \frac{\partial \varepsilon}{\partial T} \right)_{\mu} .
\end{equation}
The term in the bracket represents the sensitivity of the metallic conductivity to energy. The violation of the WF law in the longitudinal channel is therefore driven by the steepness of the DOS or the energy-dependence of the scattering rate ($\partial \tau/\partial \epsilon$), coupled with the band shift. This mechanism is maximized at band edges or Van Hove singularities where $\partial N(\epsilon)/\partial \epsilon$ is large.

\subsection{Transverse (Anomalous Hall)}\label{subb}

The derivation for the transverse transport follows a similar logical structure but stems from a different physical origin.
The intrinsic anomalous Hall conductivity arises from the Berry curvature $\Omega^z(\bm{k})$ rather than the scattering time.
We define the transverse transport kernel function $\Sigma^{xy}(\epsilon)$ as the cumulative Berry curvature up to energy $\varepsilon$:
\begin{equation}
    \Sigma^{xy}(\epsilon) = -\frac{e^2}{\hbar} \sum_n \int [d\bm{k}] \Omega^z_n(\bm{k}) \Theta(\epsilon - \epsilon_{n\bm{k}}),
\end{equation}
where $\Theta$ is the Heaviside step function.
Using integration by parts, the anomalous Hall conductivity can be cast into the familiar Sommerfeld form:
\begin{equation}
    \sigma^{xy} = \int \Sigma^{xy}(\epsilon) \left(-\frac{\partial f_0}{\partial \epsilon}\right) d\epsilon.
\end{equation}
Applying the leading-order expansion, we have $\sigma^{xy} \approx \Sigma^{xy}(\mu)$.

The thermal magnetization at the edge is given by
\begin{equation}
\bm{ M}_{Q}^{\mathrm{edge}}=-\frac{1}{\hbar} \int  [d\bm{k}]\int_{\varepsilon_{\bm{k}}}^{\infty}(\epsilon-\mu)f_0(\epsilon) \bm{\Omega}_{\bm{k}} d\epsilon ,
\end{equation}

In the presence of temperature gradient, we can write
\begin{equation}
- \bm{\nabla}_{\bm{r}}\times \bm{M}^{\text{edge}}_Q = -\bm{\nabla}T\times \frac{\partial \bm{M}_{Q}^{\text{edge}}}{\partial T}.
\end{equation}
The thermal Hall conductivity is defined as $\kappa^{xy} = -J^{\text{tr},x}_{Q}/\nabla_{y}T = {\partial {M}^{\text{edge},z}_{Q}}/{\partial T}$. Making use of \Eq{deriv}, it can be written as
\begin{equation}
\kappa^{xy} =   \frac{1}{\hbar} \int  [d\bm{k}]\int_{\varepsilon_{\bm{k}}}^{\infty}(\epsilon-\mu)\frac{\partial f_0}{\partial \varepsilon} \left( -\frac{\varepsilon - \mu}{T} + \frac{\partial \varepsilon}{\partial T} \right)  \bm{\Omega}_{\bm{k}} d\epsilon ,
\end{equation}
Following the same procedure as the longitudinal case, the transverse thermal conductivity can be expressed as:
\begin{equation}
    \kappa^{xy} = \int \frac{\Sigma^{xy}(\varepsilon)}{e^2}(\varepsilon - \mu) \left[ \frac{\varepsilon - \mu}{T} - \frac{\partial \varepsilon}{\partial T} \right] \left( - \frac{\partial f_0}{\partial \varepsilon} \right) d\varepsilon.
\end{equation}
We again separate this into the standard term $\kappa_{0}^{xy}$ and the interaction term $\kappa_{\text{Int}}^{xy}$.
The standard term yields the Wiedemann-Franz law:
\begin{equation}
    \kappa_{0}^{xy} = \frac{\pi^2 k_B^2 T}{3 e^2} \Sigma^{xy}(\mu) = L_0 T \sigma^{xy}.
\end{equation}
The deviation coefficient $\gamma(T)$ is
\begin{equation}
\gamma_{xy}(T) =  1 - \frac{1}{L_0 e} \mathcal{S}^{xy} \left( \frac{\partial \varepsilon}{\partial T} \right)_{\mu} .
\end{equation}
We now introduce the intrinsic Hall Seebeck coefficient $\mathcal{S}^{xy}$ (closely related to the anomalous Nernst effect via the Mott relation):
\begin{equation}\label{sb_xy}
    \mathcal{S}^{xy} = \frac{\pi^2 k_B^2 T}{3 e}\frac{1}{\Sigma^{xy}(\mu)} \left(\frac{\partial \Sigma^{xy}}{\partial \varepsilon}\right)_\mu.
\end{equation}
It is worth noting that unlike the longitudinal case where $\Sigma^{xx}$ depends on the density of states, $\partial \Sigma^{xy}/\partial \varepsilon$ is proportional to the Berry curvature density at the Fermi energy.
This result indicates that the violation of the WF law in the transverse channel is governed by the topological properties of the Fermi surface (via $S^{xy}$) coupled with the many-body renormalization of the band structure ($\partial \varepsilon / \partial T$).

\section{Longitudinal and Transverse Seebeck Coefficients}\label{appb}
The thermoelectric transport properties are evaluated within the framework of linear response theory. In the presence of an electric field $\mathbf{E}$ and a temperature gradient $\nabla T$, the charge current density $\mathbf{J}$ is determined by the Onsager transport coefficients according to the constitutive equation:
\begin{equation}
\mathbf{J} = \hat{\sigma} \mathbf{E} - \hat{\alpha} \nabla T,
\end{equation}
where $\hat{\sigma}$ and $\hat{\alpha}$ are the electrical conductivity and thermoelectric conductivity tensors, respectively. The Seebeck coefficient tensor $\hat{S}$, which describes the generation of a voltage gradient in response to a temperature difference under open-circuit conditions ($\mathbf{J}=0$), is defined by the relation $\mathbf{E} = \hat{S} \nabla T$. By setting $\mathbf{J}=0$ in the constitutive equation, we obtain the relationship $\hat{S} = \hat{\sigma}^{-1} \hat{\alpha}$. For a two-dimensional system with the resistivity tensor $\hat{\rho} = \hat{\sigma}^{-1}$, the components of the Seebeck tensor are given by $S_{ij} = \sum_k \rho_{ik} \alpha_{kj}$. Explicitly expanding these matrix products yields the general expressions for the longitudinal ($S_{xx}$) and transverse ($S_{xy}$) Seebeck coefficients:
\begin{equation}
S_{xx} = \rho_{xx} \alpha_{xx} - \rho_{xy} \alpha_{xy} = \frac{\sigma_{xx} \alpha_{xx} + \sigma_{xy} \alpha_{xy}}{\sigma_{xx}^2 + \sigma_{xy}^2}
\end{equation}
\begin{equation}
S_{xy} = \rho_{xx} \alpha_{xy} + \rho_{xy} \alpha_{yy} = \frac{\sigma_{xx} \alpha_{xy} - \sigma_{xy} \alpha_{yy}}{\sigma_{xx}^2 + \sigma_{xy}^2}
\end{equation}
It is important to clarify the precise physical meaning of the response tensor $\hat{S}$ introduced in Eq. (3) . While this tensor dictates the interaction-induced deviation of the Wiedemann-Franz law, its components do not strictly equate to the macroscopic Seebeck tensor measured under open-circuit conditions. For the longitudinal channel, the diagonal component of Mott ratio $\mathcal{S}_{xx} = \alpha_{xx}/\sigma_{xx}$ accurately approximates the conventional longitudinal Seebeck coefficient only in the limit of a vanishingly small Hall angle where $\sigma_{xy} \ll \sigma_{xx}$. When the transverse conductivity becomes comparable to the longitudinal one, the full tensor inversion including cross terms must be considered. For the transverse channel, the off-diagonal component $\mathcal{S}_{xy} = \alpha_{xy}/\sigma_{xy}$, rather than the true transverse Seebeck coefficient $S_{xy}$, fundamentally governs the transverse Wiedemann-Franz deviation. It directly measures the interplay between the anomalous Hall and anomalous thermoelectric responses, making it a highly sensitive probe for the temperature-dependent band renormalization.

\section{Mean Field Decomposition of the  Local Interactions}
The interacting Hamiltonian includes the noninteracting terms and the on site repulsive Hubbard interaction. To treat this many body interaction term within the Hartree Fock mean field approximation,  we express the number operators as the sum of their expectation values and fluctuations, taking $n_{i\sigma} = \langle n_{i\sigma} \rangle + \delta n_{i\sigma}$. Substituting this expression into the interaction term and neglecting the second order fluctuation terms $\mathcal{O}(\delta n^2)$, the Hubbard interaction can be approximately decoupled into a bilinear form:
\begin{equation}
H_{\text{int}}^{\text{MF}} = U n_{i\uparrow}n_{i\downarrow} \approx U(\langle n_{i\uparrow} \rangle n_{i\downarrow} + \langle n_{i\downarrow} \rangle n_{i\uparrow} - \langle n_{i\uparrow} \rangle \langle n_{i\downarrow} \rangle).
\end{equation}
To reveal a clearer physical picture of the system, we introduce the local charge density $\langle n_i \rangle = \langle n_{i\uparrow} \rangle + \langle n_{i\downarrow} \rangle$ and the local magnetization along the z direction $\langle m_i \rangle = \langle n_{i\uparrow} \rangle - \langle n_{i\downarrow} \rangle$. Utilizing these two order parameters, the spin dependent mean field potential at site $i$ can be rewritten as:
\begin{equation}
H_{\text{int}}^{\text{MF}} = \sum_{i\sigma} \left[ \frac{U}{2}\langle n_i \rangle - \sigma\frac{U}{2}\langle m_i \rangle \right] c_{i\sigma}^{\dagger}c_{i\sigma} - \frac{U}{4}\left(\langle n_i \rangle^2 - \langle m_i \rangle^2\right).
\end{equation}
In our calculations, owing to the translation symmetry we have $\langle n_i \rangle = n$ , the first term containing $\langle n_i \rangle$ acts as an overall constant energy shift and can be directly absorbed into the chemical potential. Therefore, the core physical properties of the system are primarily determined by the magnetic order parameter $\langle m_i \rangle$. This term physically acts as a site dependent local Zeeman field, which can spontaneously break time reversal symmetry and alter the band topology of the system.

\section{Numerical Self Consistency Iteration Algorithm}
Because the mean field Hamiltonian inherently depends on the expectation values of its own eigenstates $\langle m_i \rangle$ , we must solve this nonlinear problem through a rigorous numerical self consistency procedure. For the crystal lattice system, we treat the magnetizations on the two sublattices $m_A$ and $m_B$ as independent variational parameters to be iterated. The specific algorithm steps are detailed below:
\subsection{Parameter Initialization and Hamiltonian Diagonalization }
We start by assigning a specific or random initial guess for the sublattice magnetic order parameters $m_A^{(0)}$ and $m_B^{(0)}$. An appropriate initial choice helps guide the system to converge toward the correct ground state. At the kth iteration step, we use the current order parameters to construct the total mean field Hamiltonian matrix $H^{\text{MF}}_{\text{int}}(m_A^{(k)}, m_B^{(k)})$ in momentum space. We exactly diagonalize this matrix to obtain the corresponding energy eigenvalues $E_{nk}$ and the complete set of eigenvectors.

\subsection{State Update and Chemical Potential Search}
After computing the new band structure, we must reevaluate the magnetization of the system. The new local magnetization $m_{\alpha}^{\text{new} }$ is obtained by summing over all occupied states below the Fermi energy:
\begin{equation}
m_{\alpha}^{\text{new}} = \frac{1}{N_k} \sum_{k} \sum_{n \in \text{occ.}} \langle nk |\sigma_z^{\alpha} | nk \rangle.
\end{equation}
Here $N_k$ is the total number of discrete momentum points in the Brillouin zone, and $\alpha$ represents the sublattice index. During this process, to ensure the system remains in the predefined metallic state, the chemical potential $\mu$ must be adjusted dynamically at each step to maintain a strictly fixed total electron filling number, setting $n=2.1$ for instance.

\subsection{Linear Mixing and Convergence Criterion: }
 To suppress numerical oscillations common in nonlinear iterations and stabilize the convergence process , we employ a linear mixing method to generate the input parameters for the subsequent step. The input for the $k+1$ step is:
\begin{equation}
m_{\alpha}^{(k+1)} = (1-\eta)m_{\alpha}^{(k)} + \eta m_{\alpha}^{new}.
\end{equation}
where $\eta$ is the mixing parameter, set to 0.2 in our calculations. We repeat the diagonalization and linear mixing steps until the maximum absolute difference between the order parameters of two successive iterations falls below a predefined tolerance threshold. In our computational model, the convergence condition is strictly set to $|m_{\alpha}^{(k+1)} - m_{\alpha}^{(k)}| < 10^{-6}t$. Once this condition is met, the self consistency loop terminates. The converged parameters and the renormalized band structure are then utilized for the subsequent accurate evaluation of topological invariants and transport coefficients.

\end{document}